\documentclass[12pt,preprint]{aastex}
\usepackage{lineno}
\usepackage{times}

\slugcomment{Version: \today}
\linenumbers

\def\deg{\hbox{$^\circ$}}

\shorttitle{{\it Suzaku} Observations of Unidentified {\it Fermi}-LAT Sources}
\shortauthors{Maeda et al.}  

\title{Unraveling the Nature of Unidentified High Galactic Latitude {\it Fermi}/LAT Gamma-ray Sources with {\it Suzaku}}

\author{K. Maeda\altaffilmark{1}, J. Kataoka\altaffilmark{1},
T. Nakamori\altaffilmark{1}, \L . Stawarz\altaffilmark{2,\,3},
R. Makiya\altaffilmark{4}, T. Totani\altaffilmark{4}, 
C.~C. Cheung\altaffilmark{5}, D. Donato\altaffilmark{6,\,7},
N. Gehrels\altaffilmark{7}, P. Saz~Parkinson\altaffilmark{8}
Y. Kanai\altaffilmark{9}, 
N. Kawai\altaffilmark{9}, Y. Tanaka\altaffilmark{2}, 
R. Sato\altaffilmark{2}, T. Takahashi\altaffilmark{2}, and
Y. Takahashi\altaffilmark{1} 
}

\email{ko-t.maeda.x-6@ruri.waseda.jp}

\altaffiltext{1}{Research Institute for Science and Engineering, Waseda University, 3-4-1 Okubo, Shinjuku, Tokyo, 169-8555, Japan} 
\altaffiltext{2}{Institute of Space and Astronautical Science (ISAS), Japan Aerospace Exploration Agency (JAXA), 3-1-1 Yoshinodai, Chuo-ku, Sagamihara, Kanagawa, 252-5510 Japan}
\altaffiltext{3}{Astronomical Observatory, Jagiellonian University, ul. Orla 171, Krak\'ow 30-244, Poland}
\altaffiltext{4}{Department of Astronomy, Kyoto University, Kitashirakawa, Sakyo-ku, Kyoto, 606-8502, Japan}
\altaffiltext{5}{NRC Research Associate, Space Science Division, Naval Research Laboratory, Washington, DC 20375, USA} 
\altaffiltext{6}{Center for Research and Exploration in Space Science and Technology (CRESST)}
\altaffiltext{7}{NASA Goddard Space Flight Center, Greenbelt, MD 20771, USA}
\altaffiltext{8}{Santa Cruz Institute for Particle Physics (SCIPP), University of California, Santa Cruz, Natural Sciences II, Room 313, 1156 High Street, Santa Cruz, CA 95064, USA}
\altaffiltext{9}{Department of Physics, Tokyo Institute of Technology, 2-12-1, Ohokayama, Meguro, Tokyo, 152-8551, Japan}

\begin{abstract}

Here we report on the results of deep X-ray follow-up observations of
 four unidentified $\gamma$-ray sources detected by the {\it Fermi}/LAT 
 instrument at high Galactic latitudes using the X-ray Imaging
 Spectrometers on-board the {\it Suzaku} satellite. All of the studied
 objects were detected with high significance during the first 3-months
 of {\it Fermi}/LAT operation, and subsequently better localized in the
 first {\it Fermi}/LAT catalog (1FGL). For some of them, possible
 associations with pulsars and active galaxies have subsequently been
 discussed, and our observations provide an important contribution to
 this debate. In particular, a bright X-ray point source has been found within 
 the $95\%$ confidence error circle of 1FGL\,J1231.1--1410. The
 X-ray spectrum of the discovered {\it Suzaku} counterpart of
 1FGL\,J1231.1--1410 is well fitted by a blackbody with an
 additional power-law component. This supports the recently claimed
 identification of this source with a millisecond pulsar 
 PSR\,J1231--1411. For the remaining three {\it Fermi} objects, on the
 other hand, the performed X-ray observations are less conclusive. In
 the case of 1FGL\,J1311.7--3429, two bright X-ray point sources were
 found within the LAT 95\% error circle. Even though the X-ray spectral
 and variability properties for these sources were robustly assessed,
 their physical nature and relationship with the $\gamma$-ray source
 remain uncertain. Similarly, we found several weak X-ray sources in the
 field of 1FGL\,J1333.2+5056, one coinciding with the high-redshift
 blazar CLASS\,J1333+5057. We argue that the available data are
 consistent with the physical association between these two objects,
 although the large positional uncertainty of the $\gamma$-ray source
 hinders a robust identification. Finally, we have detected 
 an X-ray point source in the vicinity of 1FGL\,J2017.3+0603. This 
 {\it Fermi} object was recently suggested to be associated
 with a newly discovered millisecond radio pulsar PSR\,J2017+0603, 
 because of the spatial coincidence and the detection of the $\gamma$-ray 
 pulsations in the light curve of 1FGL\,J2017.3+0603.
 Interestingly, we have detected the X-ray counterpart of the
 high-redshift blazar CLASS\,J2017+0603, located within the error circle
 of the $\gamma$-ray source, while we were only able to determine an X-ray 
 flux upper limit at the pulsar position.
 All in all, our studies indicate that while a significant fraction of
 unidentified high Galactic latitude $\gamma$-ray sources is related to
 the pulsar and blazar phenomena, associations with other classes of
 astrophysical objects are still valid options. 

\end{abstract}


\begin{document}

\section{Introduction}
\label{sec:intro}

Observations with the EGRET instrument onboard the 
{\it Compton Gamma-Ray Observatory} (CGRO) in the 1990's opened a new
window in studying MeV--GeV emissions from both Galactic and
extragalactic objects. Despite over a decade of multi-wavelength
follow-up studies, more than $60\%$ of the $\gamma$-ray emitters
included in the 3rd EGRET catalog \citep[3EG;][]{3EG} 
are yet to be identified (that is, 170 out of 271). This is mainly
because of the relatively poor $\gamma$-ray localizations of EGRET sources
(typical $95\%$ confidence radii, $r_{\rm 95} \simeq 0.4\deg-0.7\deg$),
challenging the identification procedure especially for the objects
located within the Galactic plane, due to source confusion. In
particular, as much as $\simeq 90\%$ of the 3EG sources detected at
Galactic latitudes $|b|<10^{\circ}$ do not have robustly selected
counterparts at lower frequencies. On the other hand, most of the 3EG
sources at high Galactic latitudes have been associated with blazars
--- a sub-class of jetted active galactic nuclei (AGN) displaying strong
relativistic beaming --- in accordance with the expectation for the
extragalactic population to dominate the $\gamma$-ray sky at 
$|b| > 10^{\circ}$ \citep{LBAS}. Yet the unidentified fraction of the 
high Galactic latitude 3EG sources is still large 
\citep[$\simeq 30\%$; e.g.,][]{sowards03}. The situation is basically 
unchanged in the revised EGRET catalog \citep[EGR;][]{EGR}, 
even though the revised background modeling applied in the EGR resulted
in fewer $\gamma$-ray detections (188 sources in total, in contrast to
271 listed in 3EG); 87 out of 188 EGR entries remain unidentified. 

The unidentified low Galactic latitude $\gamma$-ray sources are expected
to be associated with local objects such as molecular clouds, supernova
remnants, massive stars, pulsars and pulsar wind nebulae, or X-ray
binaries \citep[see, e.g.,][and references therein]{gehrels99}.
Meanwhile, the population of unidentified high Galactic latitude
$\gamma$-ray sources is typically believed to be predominantly
extragalactic in origin,
although there is a suspected Galactic component as well 
\citep{ozel96}. For example, the brightest
steady source 
3EG~J1835+5918 located at $|b|>10\deg$ was proposed to be associated 
with an isolated neutron star \citep{mirabal00,reimer01}. The neutron 
star origin and its association with the $\gamma$-ray source has been 
confirmed with the discovery of a $\gamma$-ray pulsar at the position 
of 3EG~J1835+5918 with {\it Fermi}/LAT \citep{PSRJ1836,PSRJ0034}. 
Similarly, high-energy $\gamma$-ray pulsations were discovered with 
{\it Fermi} \citep{PSRJ2021} and {\it AGILE} \citep{AGILE} from 
PSR~J2021+3651 that was long considered as a likely pulsar counterpart 
of 3EG~J2021+3716 \citep{halpern08}.
On the other hand, blazar G74.87+1.22 (B\,2013+370) was claimed to be the 
most likely counterpart of the unidentified object 3EG~J2016+3657 located 
within the Galactic plane \citep{mukherjee00,halpern01}.
Other unidentified $\gamma$-rays sources were similarly investigated 
with varying success \citep[e.g.,][]{mukherjee04}.
 We note that
population studies, which could in principle shed some light on the
galactic/extragalactic origin of different classes of unidentified EGRET
sources, 
were impeded
by different level of background emission 
at different locations from the Galactic plane, and different EGRET
exposure for various parts of the sky \cite[see the discussion in][]{reimer01book}.
Also, variability studies were previously hampered by the limited
statistics and noncontinuous EGRET observations \citep{nolan03}.

With the successful launch of the {\it Fermi} Gamma-ray Space Telescope,
we now have a new opportunity to study $\gamma$-ray emission from
different types of high energy sources with much improved sensitivity
and localization capabilities than with EGRET. With its field of view
(five-times-larger than that of EGRET) covering $20\%$ of the sky at
every moment, and its improved sensitivity (by more than an order of
magnitude with respect to EGRET), the Large Area Telescope 
\citep[LAT;][]{atwood09} aboard {\it Fermi} surveys the entire sky 
each day down to a photon flux levels of $F_{>100\,{\rm MeV}} \simeq$ few 
$\times 10^{-7}$\,ph\,cm$^{-2}$\,s$^{-1}$. 
The first {\it Fermi}/LAT point source catalog (1FGL) already
surpasses EGRET with 1451 sources detected at significance levels 
$> 4\sigma$ within the $100$\,MeV\,$-100$\,GeV photon energy range
during the initial 11-month survey \citep{1FGL}.
Several high-latitude EGRET sources lacking low-frequency
counterparts were confirmed by  {\it Fermi}/LAT and associated with
previously unknown $\gamma$-ray blazars, as expected \citep{1LAC}.
Somewhat surprisingly, however, a number of $\gamma$-ray emitters at
$|b|>10\deg$ have been robustly identified by LAT with newly found
$\gamma$-ray pulsars via the detection of $\gamma$-ray pulsations 
\citep{1LGPC}. Most of these are in fact millisecond pulsars (MSPs). 
A diminishing, yet still significant population of unidentified 
{\it Fermi}/LAT objects remains, constituting as much as about $40\%$ of
all 1FGL sources. This includes more than 10 unidentified EGRET sources
at high Galactic latitudes, which are thus the best candidates 
for the persistent, or even ``steady'' $\gamma$-ray emitters over the
10-year-long period between the EGRET and {\it Fermi}/LAT epochs
(as indicated by their comparable photon fluxes in the 3EG and 1FGL catalogs).

Thus motivated, we started a new project to investigate the nature of
unidentified high Galactic latitude {\it Fermi} objects through deep
X-ray follow-up observations with the Japanese X-ray astronomy satellite
{\it Suzaku} \citep{mitsuda07}. This paper presents the results of the 
first year campaign conducted over the span of {\it Suzaku}-AO4 
(Apr 2009 -- Mar 2010), during which we have observed four steady/weakly 
variable {\it Fermi}/LAT sources from the 3-month {\it Fermi}/LAT Bright 
Source List \citep[0FGL;][]{0FGL}. These are denoted below accordingly to 
their 1FGL catalog entries as 1FGL\,J1231.1--1410,
1FGL\,J1311.7--3429, 1FGL\,J1333.2+5056, and 1FGL\,J2017.3+0603. Thanks
to the superb localization provided by the LAT, all the corresponding
$95\%$ error circles (typically $r_{\rm 95} \simeq 0.1\deg-0.2\deg$)
could be covered within the field-of-view of the {\it Suzaku} X-ray CCD
camera ``XIS''. Only in the case of 1FGL~J1333.2+5056, the 
{\it Suzaku} pointing does not cover the entire 95\% LAT error circle
since the localization error for this object did not improve
sufficiently between 1FGL and 0FGL.
Along with our Suzaku observations, systematic pulsar searches with 
radio telescopes have been performed for the {\it Fermi}/LAT unassociated sources.
These resulted
in the new discoveries of MSPs co-located with the 
two $\gamma$-ray sources included in our study (1FGL\,J1231.1--1410 and
1FGL\,J2017.3+0603). In both cases, {\it Fermi}/LAT eventually detected 
$\gamma$-ray pulsations as well, in accordance with the results 
in the radio domain \citep{ransom10,cognard10}. Our deep X-ray exposure 
discussed in the next sections supports the pulsar identification 
for at least 1FGL\,1231.1--1410, but is less conclusive in the case 
of 1FGL\,J2017.3+0603. For the other target from our list, 
1FGL\,J1333.2+5056, a tentative association with blazar CLASS\,J1333+5057 
was claimed in the LAT Bright AGN Sample \cite[LBAS;][]{LBAS}. Here we 
substantiate this possibility by presenting the broad-band spectral 
energy distribution (SED) for 1FGL\,J1333.2+5056/CLASS\,J1333+5057, 
including new {\it Suzaku} data, which is indeed typical of a flat 
spectrum radio quasar (FSRQ). 
Finally, the nature of the remaining 
source 1FGL\,J1311.7--3429 (for which no radio or $\gamma$-ray pulsations 
have been detected so far; Ransom et al. 2010) could not be revealed, despite the discovery 
of a likely X-ray counterpart. In particular, we found that the 
multiwavelength spectrum of 1FGL\,J1311.7--3429 is not consistent with 
neither a typical blazar nor pulsar spectrum.

In $\S$\,2, we describe the {\it Suzaku} X-ray follow-up observations
and the data reduction procedure. The results of the analysis are given
in $\S$\,3. The discussion and conclusions are presented in $\S$\,4 and
$\S$\,5, respectively. A standard $\Lambda$CDM cosmology with
$\Omega_{\Lambda} = 0.73$, $\Omega_{\rm M} = 0.27$, and $H_{\rm 0} =
71$\,km\,s$^{-1}$\,Mpc$^{-1}$ is assumed throughout the paper. 

\section{Observations and Analysis}
\label{sec:obs-analys}

\subsection{Observations and Data Reduction}
\label{sec:obs-reduct}

We observed four unidentified high Galactic latitude {\it Fermi}/LAT
objects with the {\it Suzaku} X-ray astronomy satellite
\citep{mitsuda07}. These are denoted in the 1FGL catalog as
1FGL\,J1231.1--1410, 1FGL\,J1311.7--3429, 1FGL\,J1333.2+5056, and
1FGL\,J2017.3+0603 \citep[see][]{1FGL}. All the sources but one
(1FGL\,J2017.3+0603) were already listed in the 3rd EGRET catalog
\citep{3EG} and their $\gamma$-ray fluxes are given in
Table~\ref{tab:3EG-1FGL}. The {\it Suzaku} observation logs are
summarized in Table\,\ref{tab:exposure}. The observations were made with
three out of four CCD cameras 
\citep[X-ray Imaging Spectrometers; XIS;][]{koyama07}, 
and a Hard X-ray Detector \citep[HXD;][]{kokubun07,takahashi07}.
One of the XIS sensors is a back-illuminated CCD (BI; XIS1), and the
other three XIS sensors are front-illuminated ones (FI; XIS0, XIS2, and
XIS3; the operation of XIS2 has been terminated in
November 2006). Since none of the studied sources have been detected
with the HXD, in this paper we focus on the analysis of only the XIS
data. The XIS was operated in the pointing source mode and in the normal
clocking mode during all the exposures. 

In the reduction and the analysis of the {\it Suzaku} data, HEADAS
software version 6.7 and a calibration database (CALDB; released on 2009
September 25th) were used. The XIS cleaned event dataset was obtained in
the combined $3 \times 3$ and $5 \times 5$ edit modes using
\texttt{xselect}. We excluded the data collected during the time and up
to 60 seconds after {\it Suzaku} was passing the South Atlantic Anomaly
(SAA). We also excluded the data corresponding to less than 5 degrees of
the angle between the Earth's limb and the pointing direction (the
Elevation Angle; ELV). Moreover, we excluded time windows during which
the spacecraft was passing through the low Cut-Off Rigidity (COR) of
below 
$6$\,GV.
Finally, we removed hot and flickering pixels 
\citep[using \texttt{sisclean};][]{day98}.
With all the aforementioned data selection criteria applied, the
resulting total effective exposures for all the observed sources are
summarized in Table~\ref{tab:exposure}.

\subsection{Analysis}
\label{sec:analys}

XIS images for each target were extracted from the two FI CCDs (XIS0,
XIS3) within the photon energy range from $0.4$ to $10$\,keV. In the
image analysis procedure, calibration sources located at the corners of
CCD chips were excluded. The images of Non X-ray Background (NXB) were
obtained from the night Earth data using \texttt{xisnxbgen}
\citep{tawa08}. Since the exposure times for the original data were
different from that of NXB, we calculated the appropriate
exposure-corrected original and NXB maps using \texttt{xisexpmapgen}
\citep{ishisaki07}. The corrected NXB images were next subtracted from
the corrected original images. In addition, we simulated flat sky images
using \texttt{xissim} \citep{ishisaki07}, and applied a vignetting
correction. All the images obtained with XIS0 and XIS3 were combined and
re-binned by a factor of 4. All the FI XIS images were in addition
smoothed by a Gaussian function with $\sigma = 0.'17$, 
and the resultant images are presented in section~\ref{sec:results}.
Note that the apparent features at the edge of these exposure corrected images are undoubtedly spurious due to low exposure in those regions.
For the further analysis,
source regions were carefully selected
around each detected X-ray sources within the error circle of a
studied $\gamma$-ray emitter. The corresponding background regions
with radius $3'$ were taken from the same XIS chips avoiding any
bright X-ray spots in the field. In all the cases, such source
regions were set to within $3'$ or $1'$ radii around the X-ray point
sources (because of the blurring due to the {\it Suzaku}/XIS Point
Spread Function; PSF), depending on the properties of each analyzed field. 
The source detection criterion was based on a signal-to-noise ratio 
which is defined, assuming a Poisson distribution, as a ratio of the excess 
events above a background to its standard deviation. Photon counts were 
derived from each source and background regions and we set the detection 
threshold at $4 \sigma$. The source positions and the corresponding errors 
were obtained by fitting a 2D Gaussian around each X-ray spot. The source 
detection results are summarized in Table~\ref{tab:srcdetect}.

The light curves were constructed for each potential X-ray counterpart
of the observed {\it Fermi} objects. Each light curve provides
net-counting rates, with the count rates of the corresponding background
region subtracted. In the timing analysis, the FI (XIS0, XIS3) and BI
(XIS1) CCD's light curves were combined using \texttt{lcmath}, and then
re-binned using \texttt{lcurve}. To assess statistical significances of
the flux variations, the $\chi^2$ test was applied to each constructed
dataset (probing a constant flux hypothesis with \texttt{lcstats}
command). Finally, the XIS spectra for each source region were
extracted, with the same corresponding background spectra as defined in
the image analysis (see above). RMF files for the detector response and
ARF files for the effective area were generated using \texttt{xisrmfgen}
and \texttt{xissimarfgen} \citep{ishisaki07}. In this spectral analysis,
all the selected data from the FI CCDs were co-added (using
\texttt{mathpha}) without calculating Poisson errors, and the response
files were combined with the \texttt{marfrmf} and \texttt{addrmf}
commands. Since all the studied {\it Fermi}/LAT objects are located at
high Galactic latitudes, the absorption of soft X-ray photons was set to
the Galactic one with the equivalent column density of a neutral
hydrogen, $N_{\rm H}$, as given in \citet{Colden}.
In some cases where apparent systematic features are visible as 
trends of the residuals with energy (see Figure~\ref{fig:specJ1231}), we 
attempted to use an inter-calibration constant between the FI and BI 
CCDs to improve the fits. From this inspection, we found negligible 
improvement of the fits thus we conclude that the limited photon 
statistics is the predominantly responsible for the somewhat 
unsatisfactory model fits to the data.

\section{Results}
\label{sec:results}

\subsection{1FGL\,J1231.4--1410} 
\label{subsec:J1231}

\begin{figure}[m]
\begin{center}
\includegraphics[width=150mm]{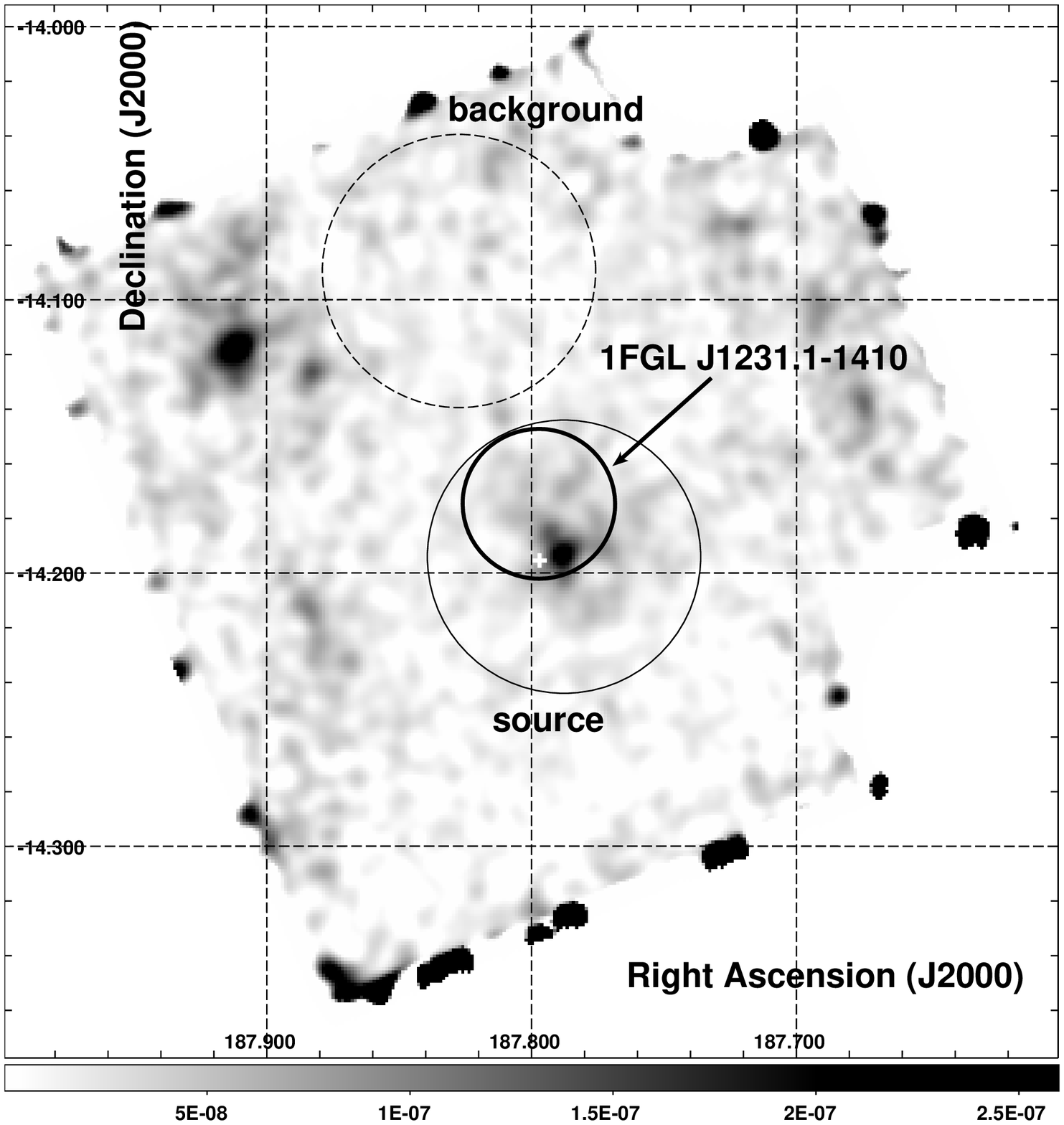} 
\end{center}
\caption{{\it Suzaku}/XIS FI (XIS0+3) image 
 of 1FGL\,J1231.1$-$1410 region in the $0.4-10$\,keV photon energy range. 
 The image shows the relative excess of smoothed photon counts
 (arbitrary units indicated in the bottom bar) and is displayed with
 linear scaling.
 The areas enclosed by solid and dashed circles are source and
 background regions, respectively. Thick solid circle denotes $95\%$
 position error of 1FGL\,J1231.1$-$1410. White cross marks the position
 of radio MSP PSR\,J1231$-$1411 \citep{ransom10}.
} 
\label{fig:imgJ1231}
\end{figure}

\begin{figure}[m]
\begin{center}
\includegraphics[width=90mm,angle=-90]{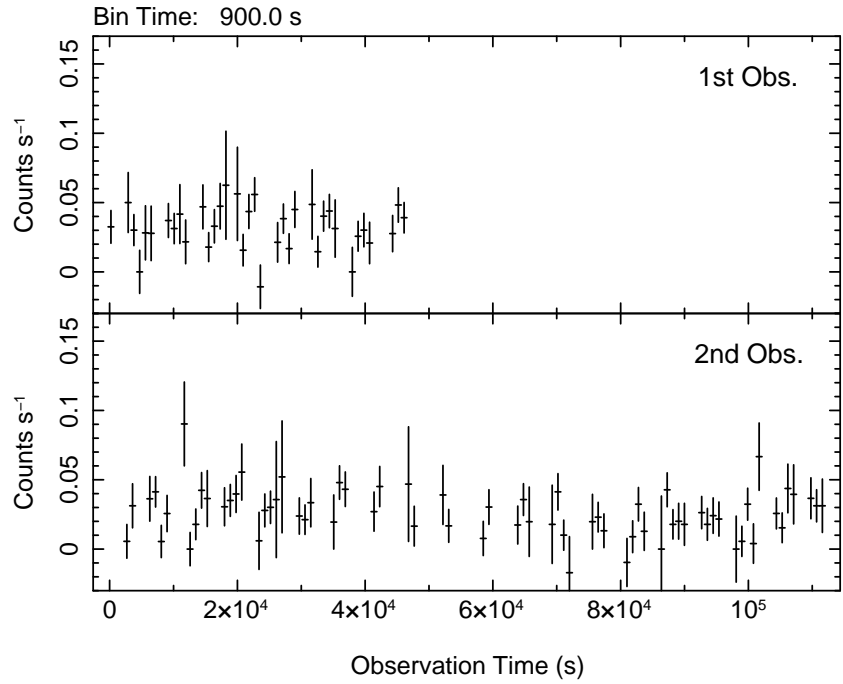}
\end{center}
\caption{{\it Suzaku}/XIS light curves of the X-ray counterpart of
 1FGL\,J1231.1$-$1410 during the 1st and the 2nd observations (upper and
 lower panels, respectively). Binning time applied is $900$\,s.
 The zero point of the upper and lower panels are MJD
 55020.9971 and 55040.2343 (TDB: Barycentric Dynamical Time).
} 
\label{fig:lcJ1231}
\end{figure}

Our {\it Suzaku} observations (interrupted for $\simeq 20$
days\footnote{The exposure was interrupted because of the Target of
Opportunity observation of GRB\,090708.}) revealed one X-ray point source 
(RA, Dec)\,$=$\,(187.\deg790(1), $-$14.\deg192(1)) within the LAT error circle 
of 1FGL\,J1231.4--1410. Figure~\ref{fig:imgJ1231} shows the corresponding 
X-ray image, prepared as described in $\S$~\ref{sec:analys}. For 
further analysis, the source extraction region was set to within a
$3'$ radius around the X-ray intensity peak, and the corresponding
background region was chosen as indicated in
Figure~\ref{fig:imgJ1231}. The light curve of the X-ray source with a
time bin of $900$\,s is presented in Figure~\ref{fig:lcJ1231}. The upper
panel shows the count rate variation during the 1st observation, while
the bottom panel shows that of the 2nd observation. The light curves of
the two periods can both be well fitted by a constant count rate of
$3.03\times10^{-2}$\,ct\,s$^{-1}$, with 
$\chi^2/$d.o.f.\,$=58.3/107$. This indicates that the X-ray emission of
the analyzed source is steady, with the $\chi^2$ probability for a
constant flux $>99\%$.  

\begin{figure}[m]
\begin{minipage}{0.5\hsize}
\begin{center}
{\small (a) Blackbody Model}
\includegraphics[width=60mm,angle=-90]{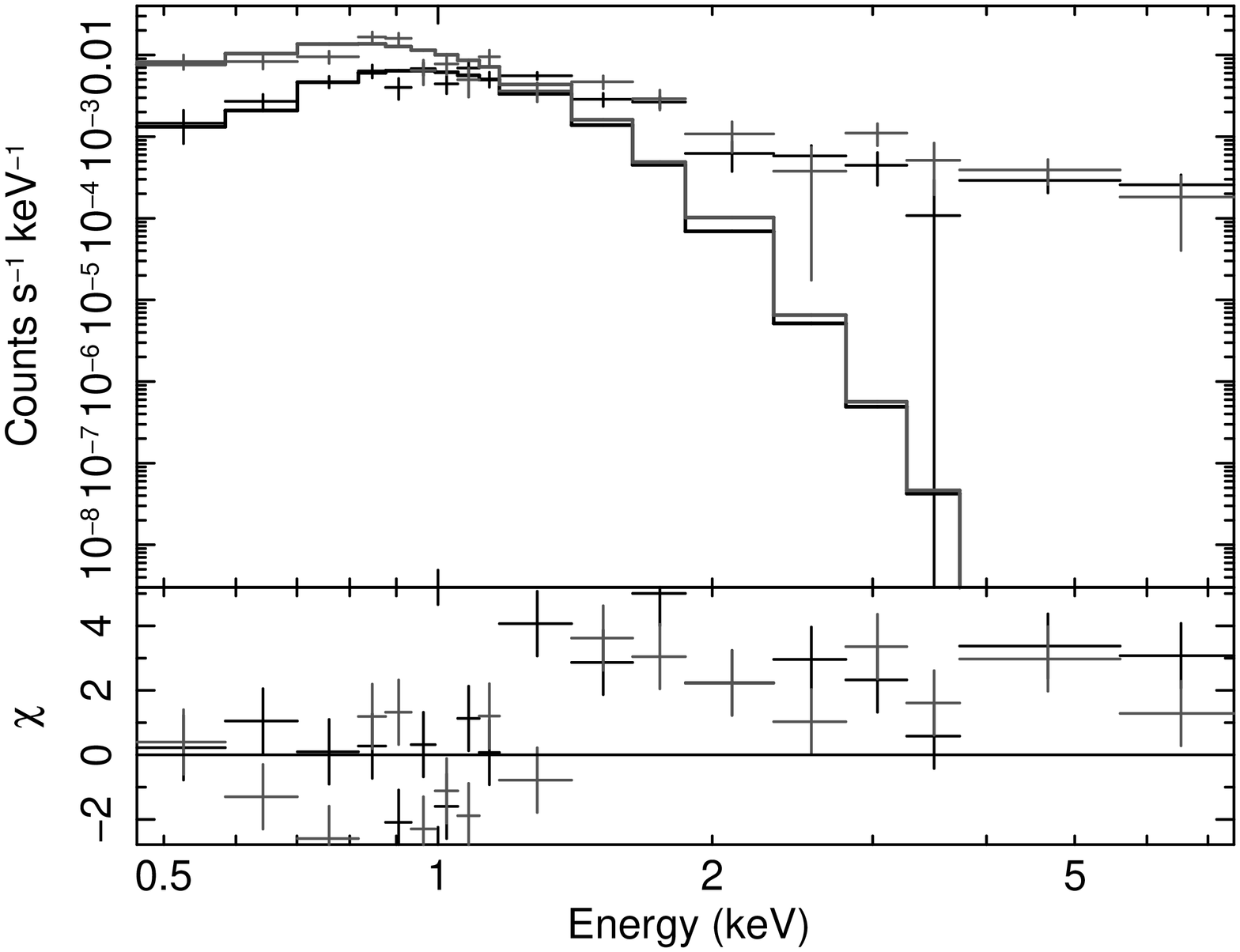} 
\end{center}
\end{minipage}
\begin{minipage}{0.5\hsize}
\begin{center}
{\small (b) Blackbody$+$Power-Law Model}
\includegraphics[width=60mm,angle=-90]{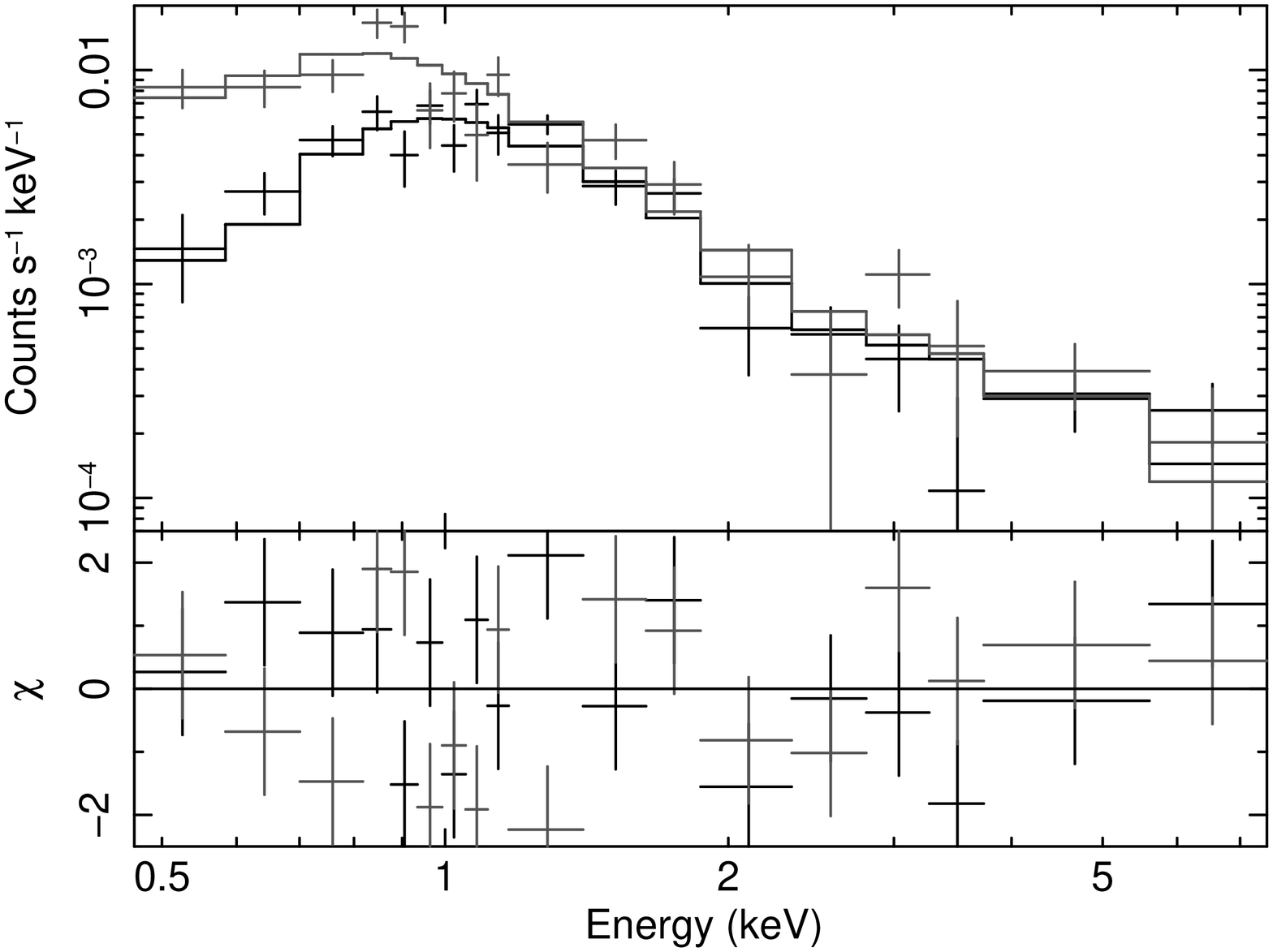} 
\end{center}
\end{minipage}
\caption{{\it Suzaku}/XIS spectra of the X-ray counterpart of
 1FGL\,J1231.1$-$1410 in the photon energy range $0.4-7.0$\,keV fitted
 with the blackbody model (a) and blackbody$+$power-law model (b). FI data
 are shown in black, and BI data in gray.} 
\label{fig:specJ1231}
\end{figure}

The X-ray spectrum of the {\it Suzaku} source, which we propose to be
the most likely counterpart of 1FGL\,J1231.1--1410, 
is shown in Figure~\ref{fig:specJ1231}.
The energy range used for the fitting was set as $0.4-7.0$\,keV. First,
we fit the X-ray spectrum by a blackbody emission moderated by the
Galactic absorption only \citep{morrison83}.
This fit was not acceptable, however, due to significant residuals above
$2$\,keV ($\chi^2/$d.o.f.\,$=128.1/34$, see Figure\,\ref{fig:specJ1231},
left panel, where the excess emission above $2$\,keV has been enhanced
by fixing the black body parameters to those determined by the data
below $2$\,keV only). The situation was essentially unchanged when the
column density was treated as a free parameter. In this
case, the residuals above $2$\,keV remained, but the returned 
value of $N_{\rm H}$ was then consistent with zero. 
To account for the $>2$\,keV emission, we therefore added a
power-law component to the thermal one, and fixed $N_{\rm H} = 0$. The
goodness of the fit was in this way substantially improved to $\chi^2$
of $55.46/32$, supporting the presence of a non-thermal tail in the
spectrum of the analyzed object (see Figure~\ref{fig:specJ1231}). In
order to further confirm the reality of this finding, we analyzed the
highest quality FI CCD (XIS0, XIS3) data which had sufficient photon
statistics within the $2-8$\,keV range, examining various approaches for
the background estimation, namely (i) the background taken from the same
CCD chips but off-axis, as given in Figure~\ref{fig:imgJ1231}, (ii)
the concentric ring background surrounding the source region on the same
CCD chips, and (iii) the background for the same region as the source
estimated from the Lockman Hole observation taken with the same XIS mode
at nearby dates (OBS ID = 104002010). In all of the examined approaches
the presence of the non-thermal component in the X-ray spectrum of
1FGL\,J1231.1--1410 could be confirmed at high significance, as
summarized in Tables~\ref{tab:specfitJ1231} and \ref{tab:ftest}. 

To sum up, we conclude that the X-ray counterpart of 1FGL\,J1231.1--1410
is robustly characterized by a blackbody-type spectrum with a
temperature of $kT \simeq 0.16 \pm 0.03$\,keV plus a power-law tail with
the photon index of $\Gamma \simeq 1.79^{+0.40}_{-0.17}$. The energy
flux of the non-thermal component is 
$S_{2-8\,{\rm keV}}\simeq5.81\times10^{-14}$\,erg\,cm$^{-2}$\,s$^{-1}$,
which can be compared with the {\it Fermi}/LAT energy flux 
$S_{0.1-10\,{\rm GeV}}\simeq(1.06\pm0.06)\times10^{-10}$\,erg\,cm$^{-2}$\,s$^{-1}$, 
as given in the 1FGL catalog. Thus, the extrapolation of the X-ray
power-law component to the $\gamma$-ray range yielding the $0.1-10$\,GeV
energy flux $\simeq 5.74 \times 10^{-13}$\,erg\,cm$^{-2}$\,s$^{-1}$,
falls orders of magnitudes below the observed one. This implies either a 
multi-component character or a concave spectral form of the high-energy
X-ray--to--$\gamma$-ray continuum of 1FGL\,J1231.1--1410, and both
possibilities should be kept in mind in the context of a very likely
association of the discussed source with a MSP.
Indeed, the MSP PSR\,J1231--1411 (marked by a white cross in
Figure~\ref{fig:imgJ1231}) was recently found by \citet{ransom10}
via the detection of radio pulsations with the pulse period of $3.68$\,ms
within the LAT error circle of 1FGL\,J1231.1--1410 using the Green Bank
Telescope (GBT), just after our {\it Suzaku} observations. 
In addition, the {\it Fermi} spectrum shows a cut-off 
at around a few GeV, which is consistent with the typical spectrum of 
MSPs \citep{ransom10}.
The X-ray emitter observed by {\it Suzaku} is located roughly $40''$
away from the newly discovered MSP PSR\,J1231--1411 \citep[][see
Figure~\ref{fig:imgJ1231}]{ransom10}, but considering the limited
pointing accuracy of the {\it Suzaku}/XIS ($\lesssim 1'$), both objects 
can be considered as co-spatial. In fact, as described in \citet{ransom10},
a {\it Swift}/XRT source at (RA, Dec)\,$=$\,(187.7972,$-$14.1953)
coinciding with the Suzaku one, was found to be positionally
consistent (within the 90\% error of $5.''5$) with that
of the MSP PSR\,J1231--1411.

\subsection{1FGL\,J1311.7--3429}
\label{subsec:J1311}

\begin{figure}[m]
\begin{center}
\includegraphics[width=150mm]{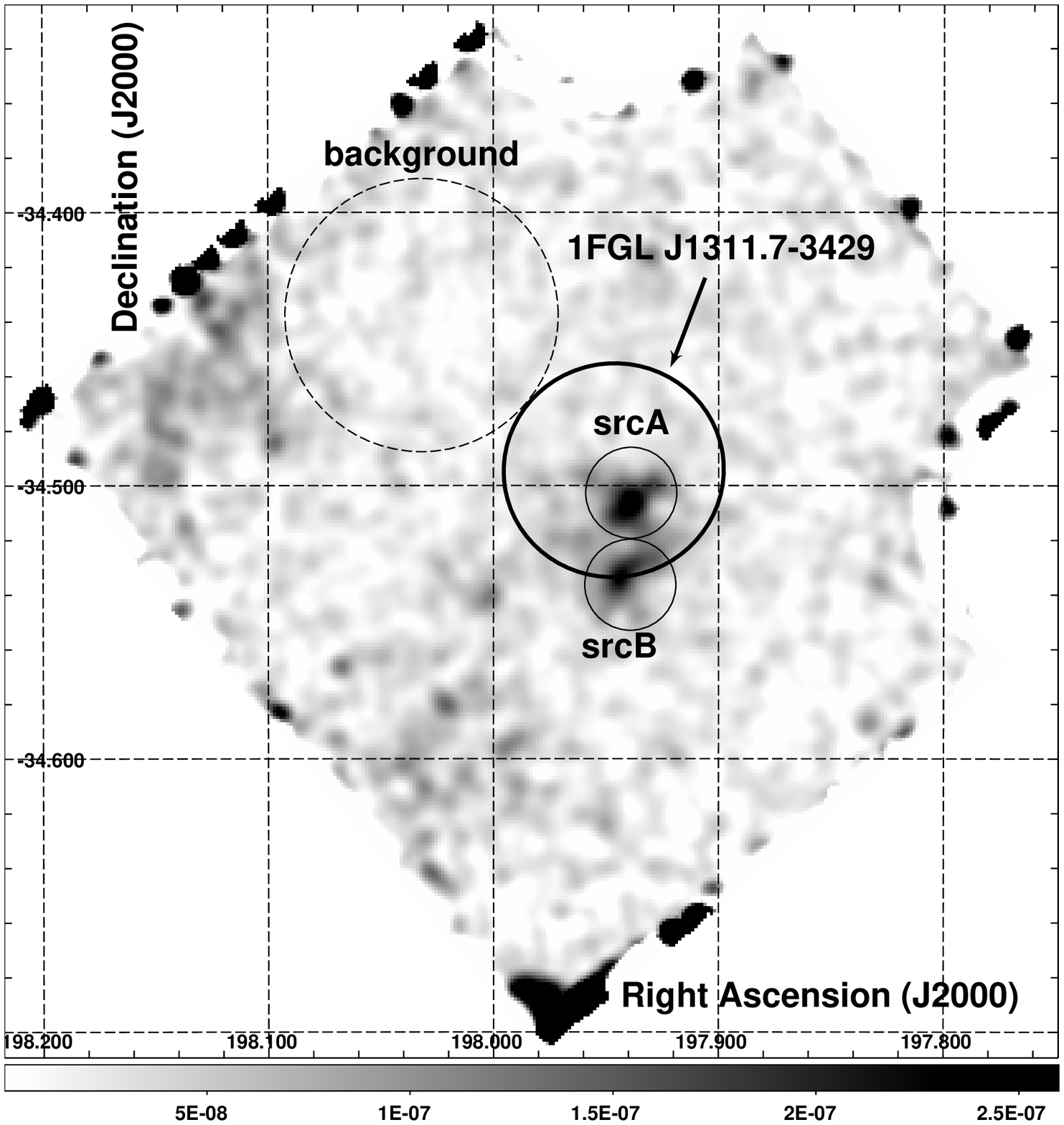} 
\end{center}
\caption{{\it Suzaku}/XIS FI (XIS0+3) image 
 of 1FGL\,J1311.7$-$3429 region in the $0.4-10$\,keV photon energy
 range. 
 The image shows the relative excess of smoothed photon counts
 (arbitrary units indicated in the bottom bar) and is displayed with
 linear scaling.
 The regions enclosed by solid and dashed circles are source and
 background regions, respectively.
 Thick solid circle denotes $95\%$ position error of 
 1FGL\,J1311.7$-$3429. Within this error circle, two potential X-ray
 counterparts of the $\gamma$-ray source were found: src\,A and src\,B.}
\label{fig:imgJ1311}
\end{figure}

\begin{figure}[m]
\begin{center}
\includegraphics[width=90mm,angle=-90]{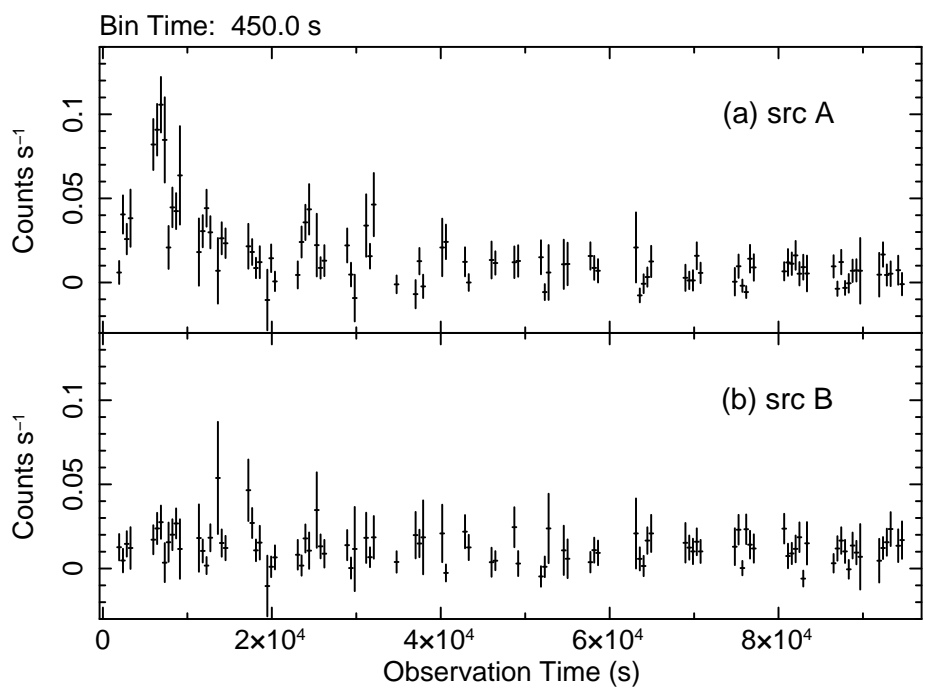} 
\end{center}
\caption{{\it Suzaku}/XIS light curves of two potential X-ray
 counterparts of 1FGL\,J1311.7$-$3429 with 450\,s binning. The northern
 source src\,A (upper panel) showed highly significant X-ray flare in
 the first $20$\,ks of observation, during which the count rate
 increased by a factor of 10. The southern source src\,B (lower panel)
 was steady during the whole exposure.
 The zero point of src\,A and src\,B is MJD 55047.2280 (TDB).
} 
\label{fig:lcJ1311}
\end{figure}

Two X-ray point sources were found within the LAT error circle of
1FGL\,J1311.7--3429. Figure~\ref{fig:imgJ1311} shows the corresponding
X-ray image with the northern {\it Suzaku} object, src\,A, located at 
(RA, Dec)\,$=$\,(197.\deg939(1), $-$34.\deg508(2)) and the southern source, 
src\,B, positioned at (RA, Dec)\,$=$\,(197.\deg942(1), $-$34.\deg534(2)). 
Note that src\,B is situated just marginally within the edge of the 
{\it Fermi}/LAT error circle. For the further analysis, we set the source 
regions to within $1'$ radii around the respective X-ray flux maxima. 
The derived light curves of src\,A and src\,B with time bins of $450$\,s 
are presented in Figure~\ref{fig:lcJ1311} (upper and lower panels, 
respectively). As shown, during the first $20$\,ksec of the observation, 
src\,A exhibited a very rapid X-ray flare, with the count rate changing 
by a factor of 10. After the flare, however, the X-ray flux of src\,A 
remained steady. A constant fit to the light curve of src\,A returns
$\chi^2/$d.o.f.\,$=403.9/97$, and hence the variability of this source
is statistically significant. On the other hand, src\,B was
characterized by a constant flux over the duration of the exposure
($\chi^2/$d.o.f.\,$= 45.0/97$) with a count rate of
$1.3 \times 10^{-2}$\,ct\,s$^{-1}$. 

\begin{figure}[m]
\begin{minipage}{0.5\hsize}
\begin{center}
{\small (a) src\,A}
\includegraphics[width=60mm,angle=-90]{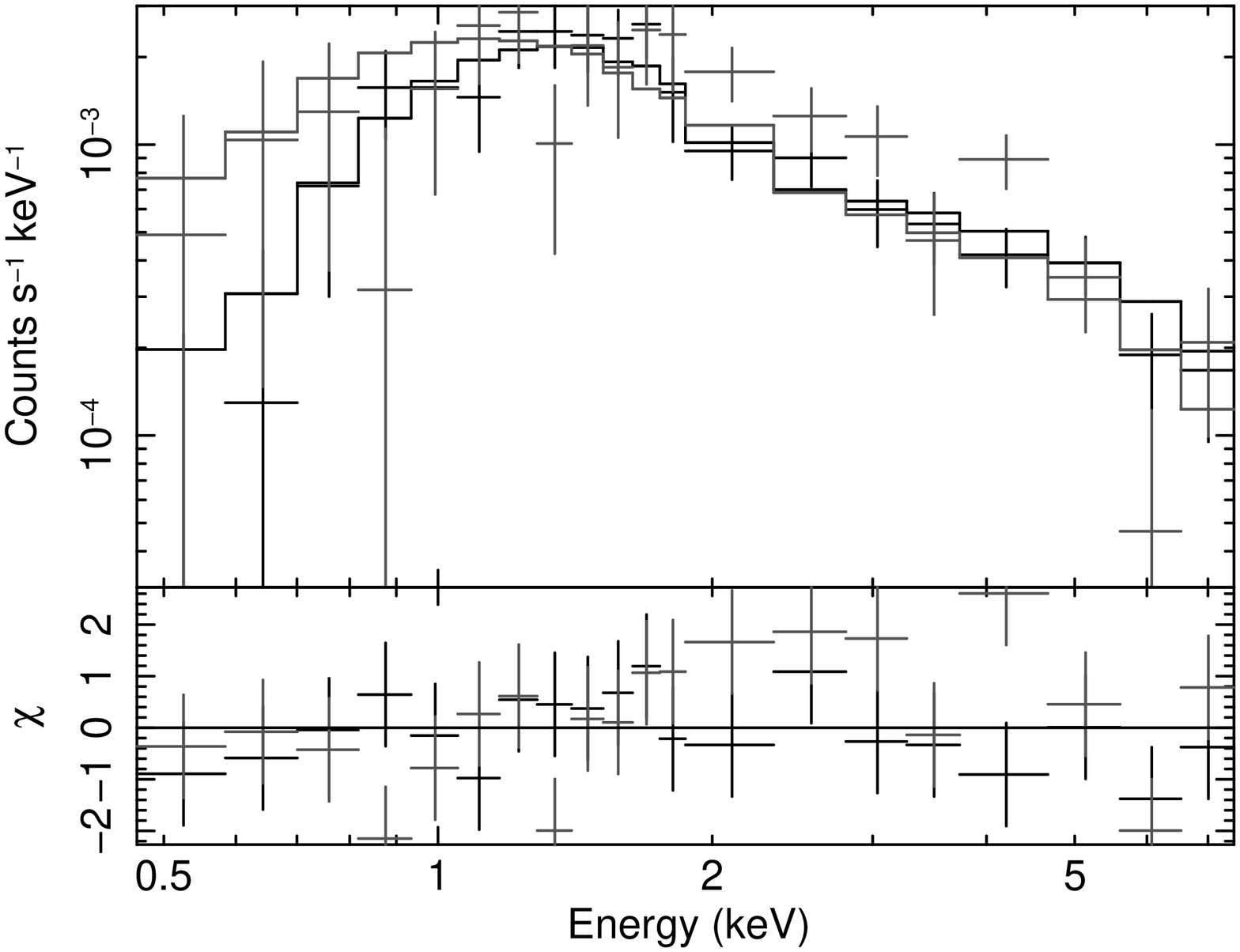} 
\end{center}
\end{minipage}
\begin{minipage}{0.5\hsize}
\begin{center}
{\small (b) src\,B}
\includegraphics[width=60mm,angle=-90]{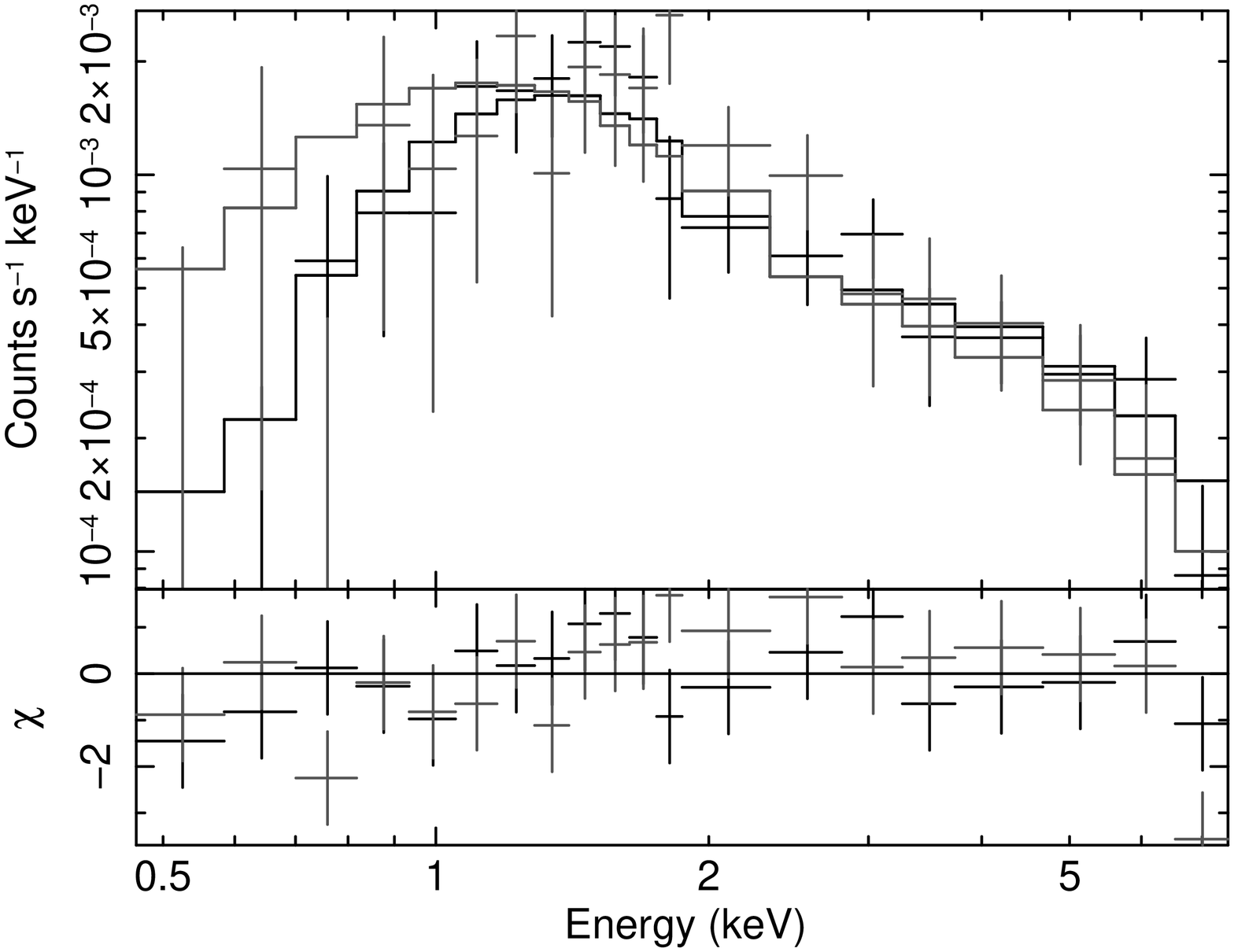} 
\end{center}
\end{minipage}
\caption{{\it Suzaku}/XIS Spectra of two possible X-ray counterparts of
 1FGL\,J1311.7$-$3429 in the photon energy range $0.4-8.0$\,keV fitted
 with the best fit power-law model. FI data are represented in black,
 and BI data in gray.}  
\label{fig:specJ1311}
\end{figure}

Figure~\ref{fig:specJ1311} shows the spectra of src\,A and src\,B within
the energy range $0.4-8.0$\,keV. The best model fits for both newly
discovered X-ray objects consist of power-law continua with photon
indices $\Gamma \simeq 1.38 \pm 0.13$ (src\,A) and 
$\Gamma \simeq 1.34\pm0.16$ (src\,B), moderated by the Galactic
absorption. The detail of the model fitting are summarized in
Table~\ref{tab:specfitJ1311}. 
Note that the observed X-ray spectra
of the two sources are very similar, and the X-ray fluxes of the objects 
are almost identical. It is important to emphasize at this point that
because of the relatively large PSF of {\it Suzaku}/XIS (a half power
diameter of $\sim3'$), it is quite difficult to separate completely
src\,A and src\,B --- located only $1.'6$ apart --- for the purpose of
the spectral analysis. 
As a result, even though it is clear we
are dealing with two physically distinct X-ray sources (each detected at high 
significance), their spectral parameters cannot be accessed robustly.

\subsection{1FGL\,J1333.2+5056}
\label{subsec:J1333}

\begin{figure}[m]
\begin{center}
\includegraphics[width=150mm]{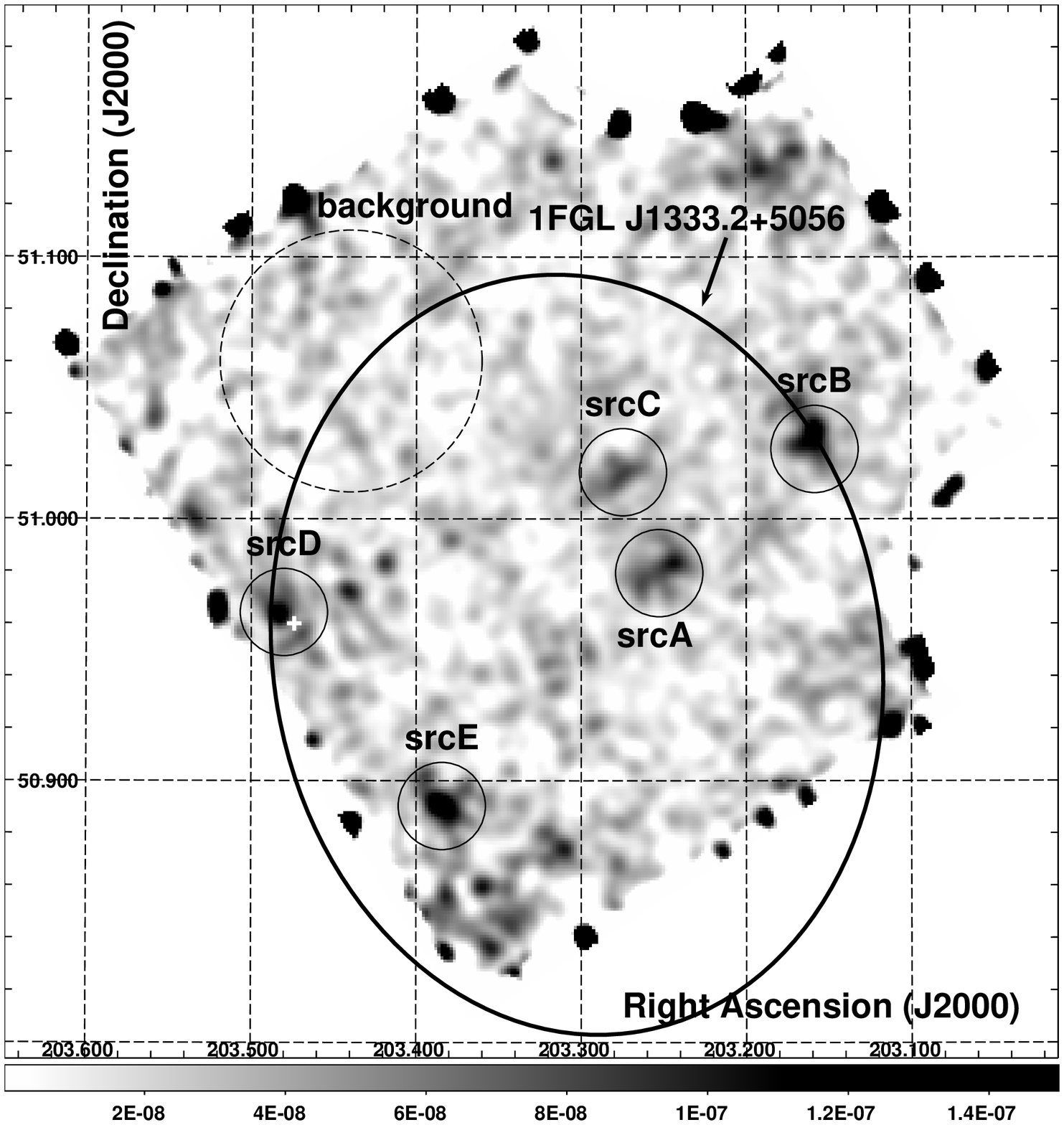} 
\end{center}
\caption{{\it Suzaku}/XIS FI (XIS0+3) image 
 of the 1FGL\,J1333.2$+$5056 region in the $0.4-10$\,keV photon energy
 range. 
 The image shows the relative excess of smoothed photon
 counts (arbitrary units indicated in the bottom bar) and is displayed
 with linear scaling. 
 The regions enclosed by solid and dashed circles are source and
 background regions, respectively. Thick solid ellipsis denotes $95\%$
 position error of 1FGL\,J1333.2$+$5056. Within this error circle,
 several potential X-ray counterparts of the $\gamma$-ray object were
 found. White cross marks the position of active galaxy CLASS
 J1333$+$5057.}
\label{fig:imgJ1333}
\end{figure}

Our {\it Suzaku} observations revealed multiple regions of enhanced 
X-ray emission inside the LAT error circle of 1FGL\,J1333.2+5056, as
shown in the corresponding X-ray image in Figure~\ref{fig:imgJ1333}. The
associations of these faint X-ray sources with 1FGL\,J1333.2+5056 are
therefore quite ambiguous. Within the {\it Fermi}/LAT error circle
covered by the XIS exposure\footnote{Note that the 1FGL localization error 
for the analyzed $\gamma$-ray object did not improve sufficiently between 
0FGL and 1FGL. For this reason, we could not cover the entire 95\% LAT error 
circle of 1FGL\,J1333.2+5056 within one pointing of {\it Suzaku}/XIS.}, 
five X-ray enhancements have been found with detection significances
of more than $4\sigma$, and these are denoted here as src\,A, B, C, D
and E (see Figure~\ref{fig:imgJ1333} 
and Table~\ref{tab:srcdetect}).

\begin{figure}[m]
\begin{center}
\includegraphics[width=90mm,angle=-90]{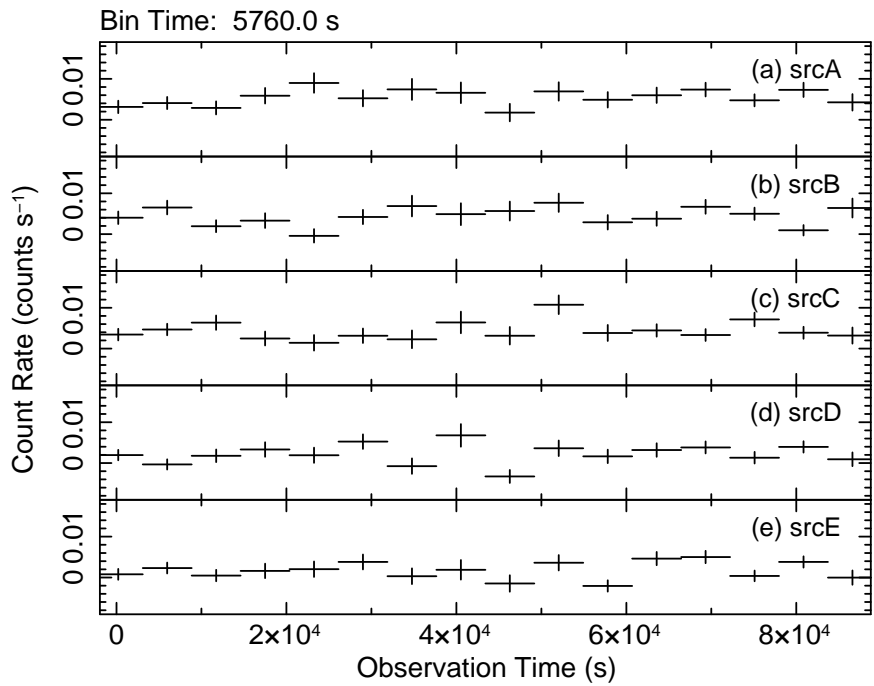} 
\end{center}
\caption{{\it Suzaku}/XIS light curves of potential X-ray counterparts
 of 1FGL\,J1333.2$+$5056 with the applied time binning of $5760$\,s.
 The zero point of time is MJD 54983.4274 (TDB).
} 
\label{fig:lcJ1333}
\end{figure}

\begin{figure}[m]
\begin{minipage}{0.5\hsize}
\begin{center}
{\small (a) src\,A \\}
\includegraphics[width=60mm,angle=-90]{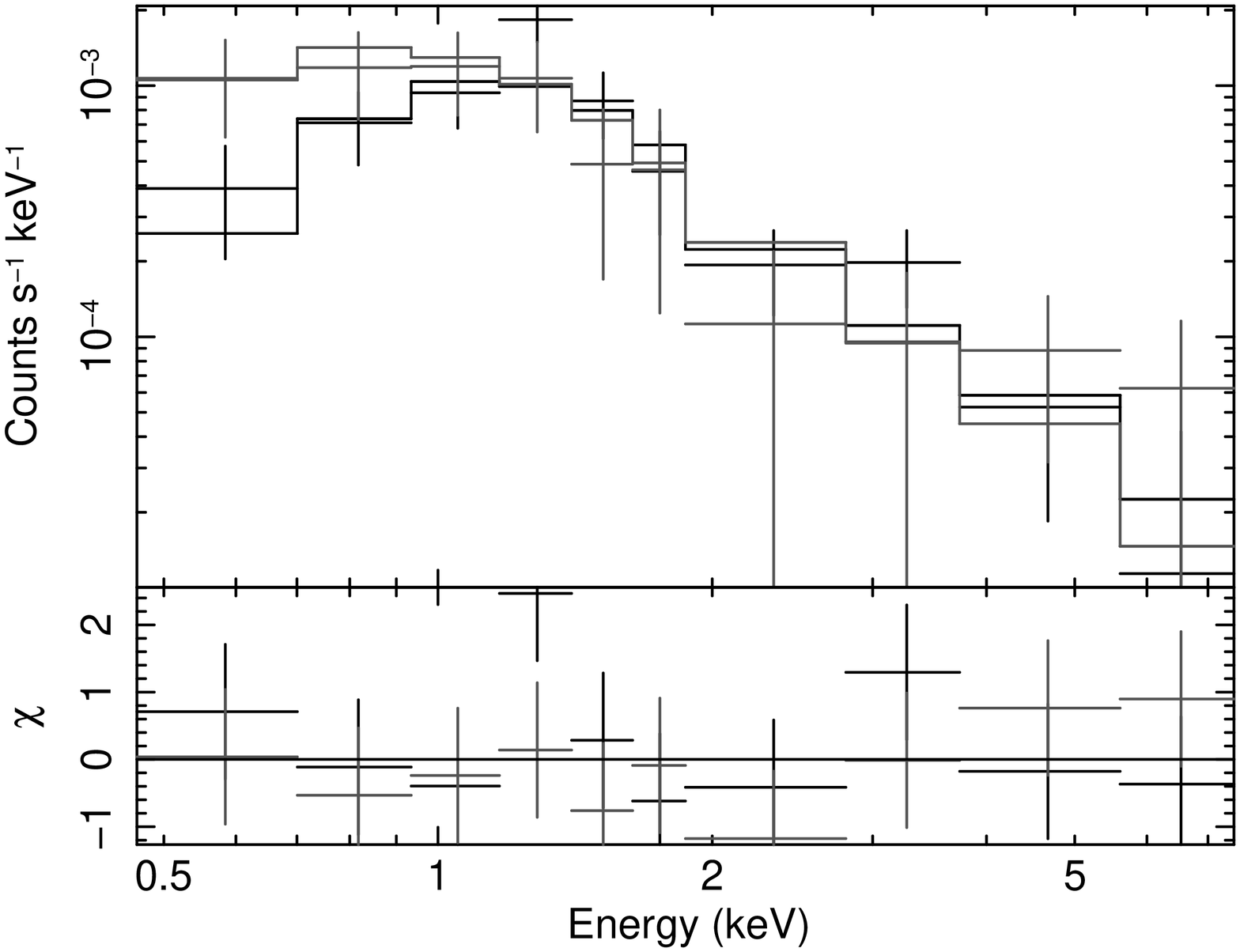} 
\vspace*{0.1cm}
\end{center}
\end{minipage}
\begin{minipage}{0.5\hsize}
\begin{center}
{\small (b) src\,B \\}
\includegraphics[width=60mm,angle=-90]{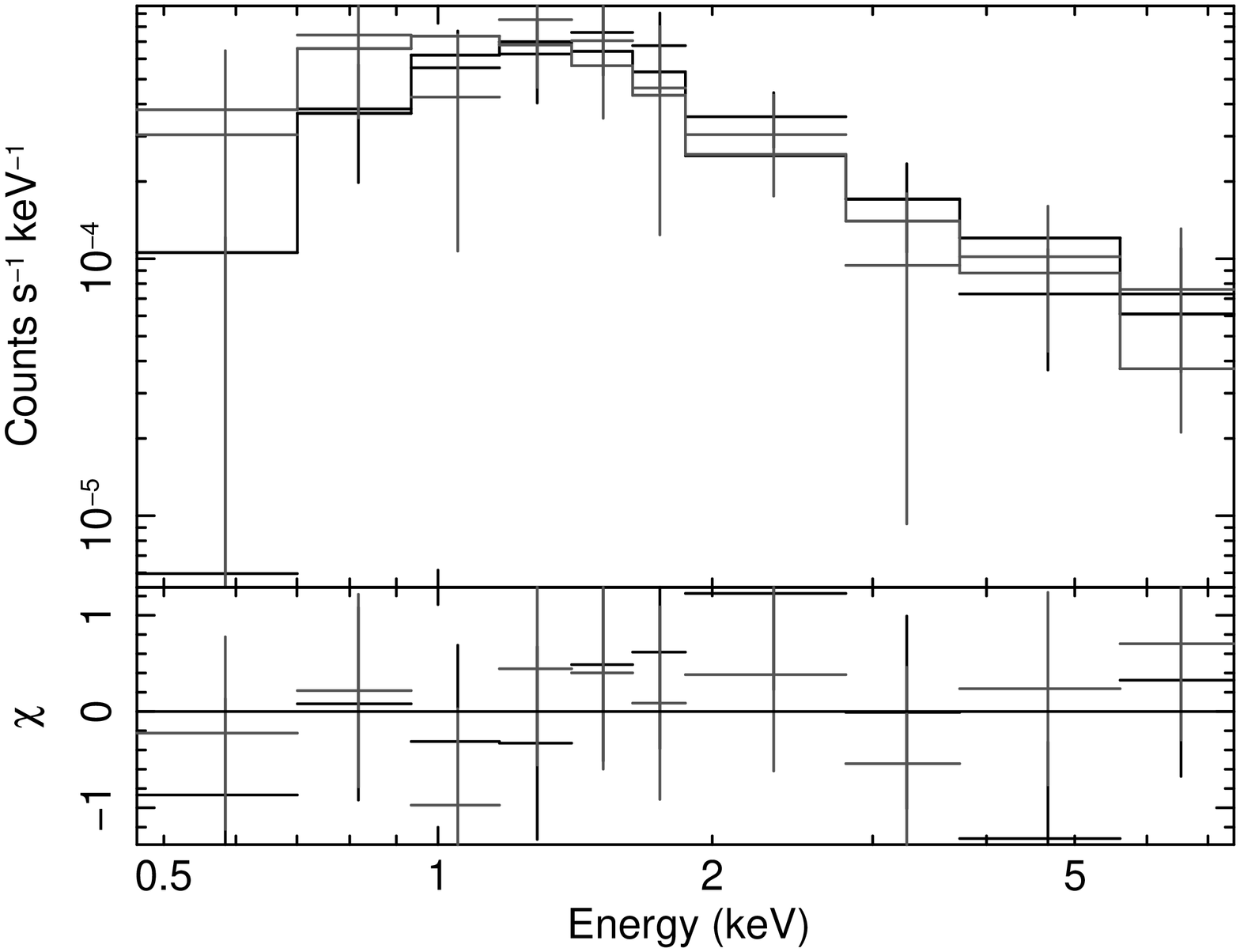} 
\vspace*{0.1cm}
\end{center}
\end{minipage}
\begin{minipage}{0.5\hsize}
\begin{center}
{\small (c) src\,C \\}
\includegraphics[width=60mm,angle=-90]{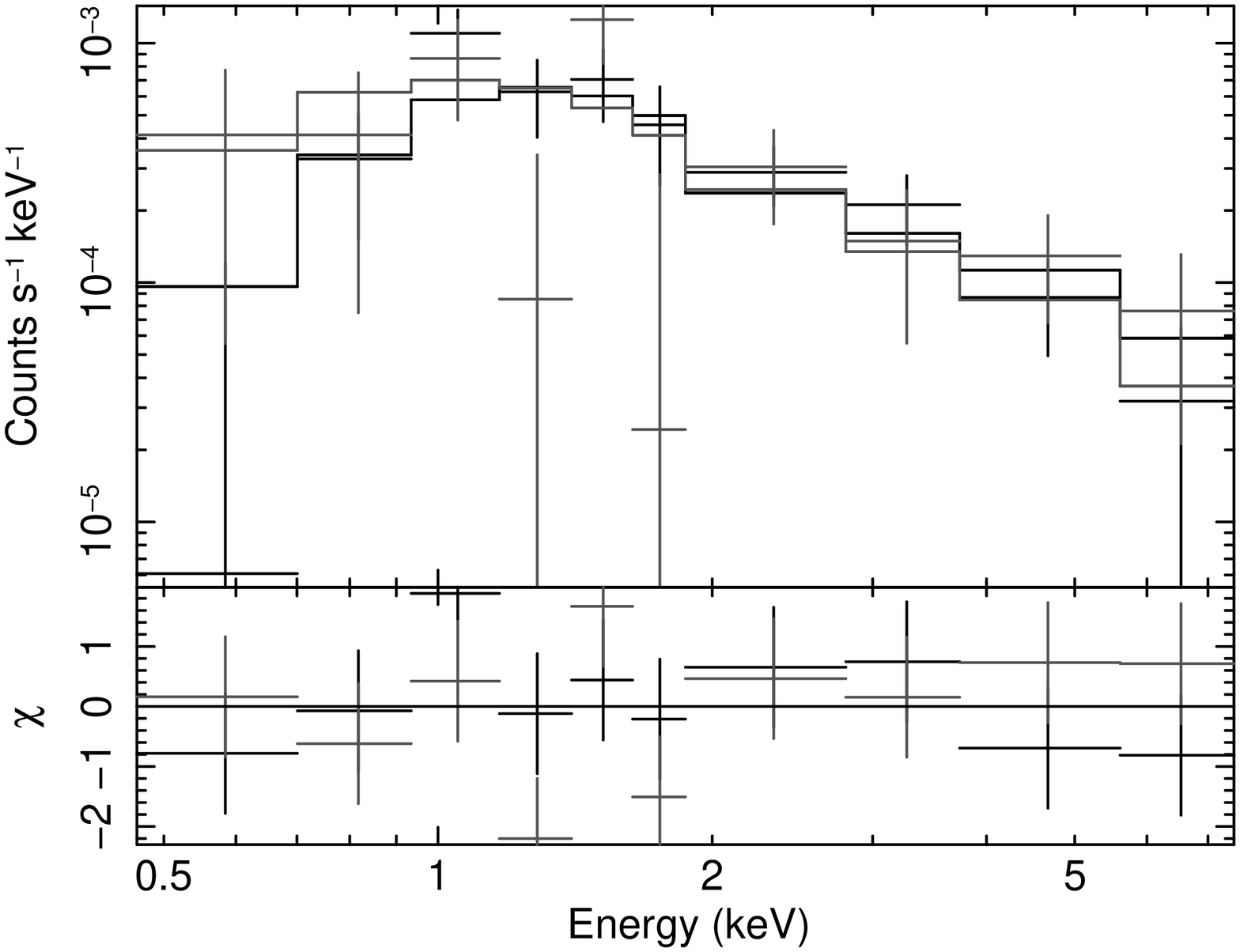} 
\vspace*{0.1cm}
\end{center}
\end{minipage}
\begin{minipage}{0.5\hsize}
\begin{center}
{\small (d) src\,D \\}
\includegraphics[width=60mm,angle=-90]{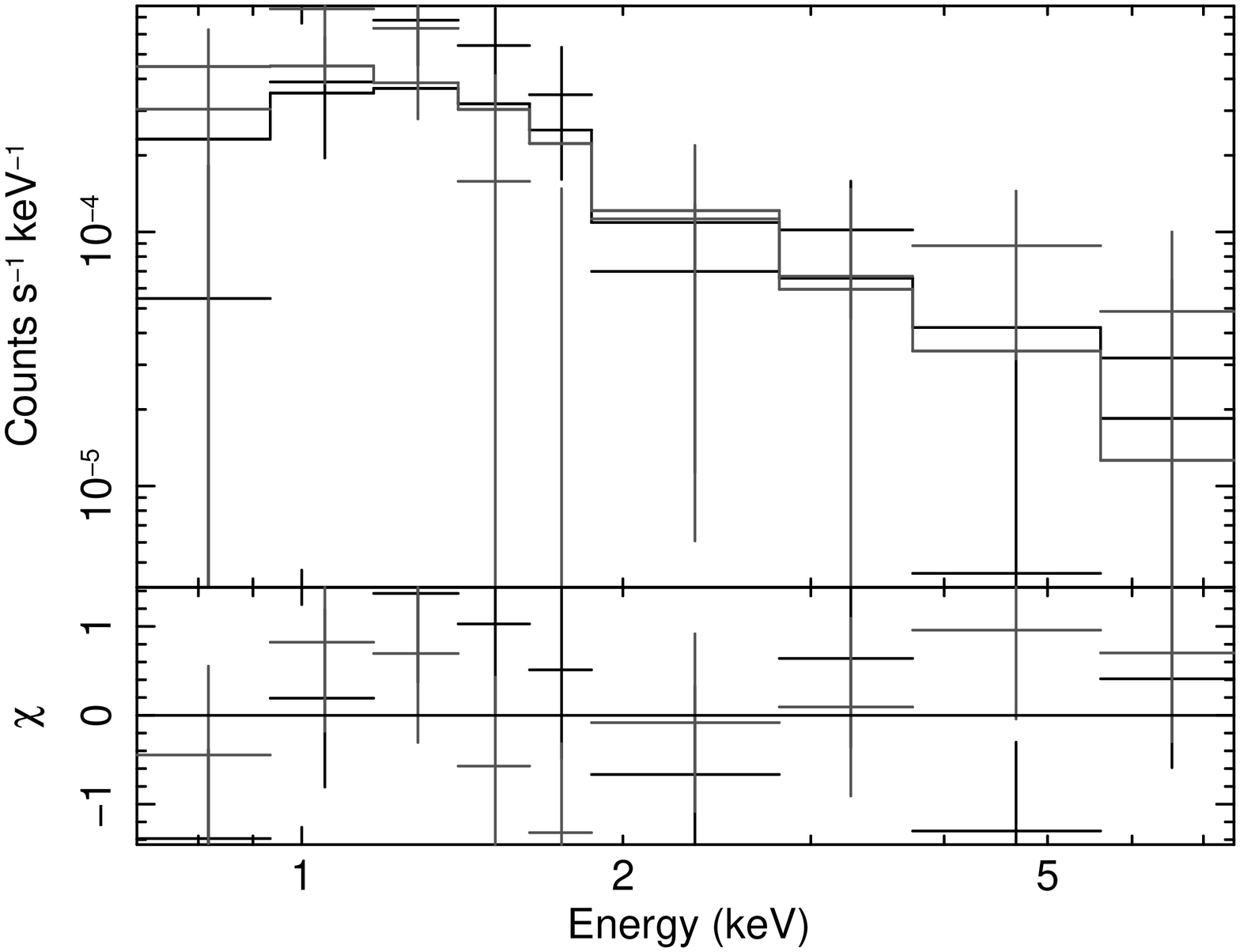} 
\vspace*{0.1cm}
\end{center}
\end{minipage}
\begin{minipage}{0.5\hsize}
\begin{center}
{\small (e) src\,E \\}
\includegraphics[width=60mm,angle=-90]{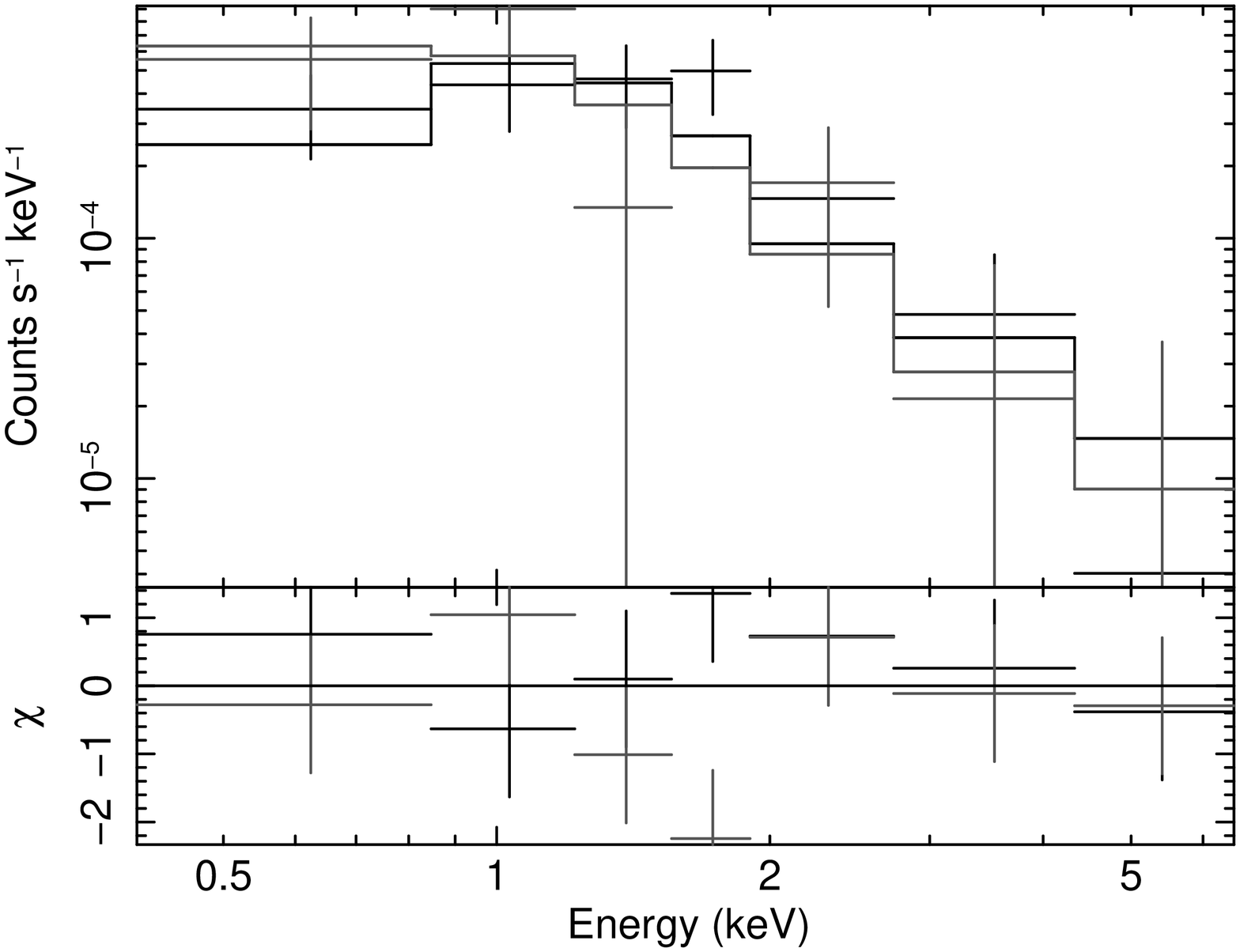} 
\end{center}
\end{minipage}
\caption{{\it Suzaku}/XIS spectra of the selected possible X-ray
 counterparts of 1FGL\,J1333.2+5056 fitted with a power-law model. FI
 data are represented in black, and BI data in gray.}   
\label{fig:specJ1333}
\end{figure}

The light curves of src\,A, B, C, D and E with $5760$\,s binning are
shown in Figure~\ref{fig:lcJ1333} in the descending order. As 
noted above, all the analyzed X-ray sources are very dim, with
X-ray fluxes at the level of $\sim10^{-14}$\,erg\,cm$^{-2}$\,s$^{-1}$. 
Hence, we could not 
assess 
the variability properties of the selected objects by means of the $\chi^2$
test with a constant flux hypothesis (see
Table~\ref{tab:lcfitJ1333}). The spectra of the five X-ray sources, all
extracted within $1'$ source radii, are shown in
Figure~\ref{fig:specJ1333}. Again, limited photon statistics precluded
any detailed analysis, and therefore in the model fitting we applied
only single power-law models moderated by the Galactic absorption. The
results are summarized in Table~\ref{tab:specfitJ1333}. 
We also emphasize that the 1FGL error circle unfortunately runs off the 
edge of {\it Suzaku} field of view. For all these reasons, we cannot 
persuasively identify an X-ray counterpart of the $\gamma$-ray source
1FGL\,J1333.2+5056. Nevertheless, we note that one of the X-ray enhancements, 
src\,D, coincides
with the $z=1.362$ FSRQ CLASS\,J1333+5057 
\citep[marked in Figure~\ref{fig:imgJ1333} by a white cross;][]{shaw09}, 
listed in the 1FGL as 
a possible
association with 1FGL\,J1333.2+5056.
Note however a relatively low significance of the detection 
of this source with {\it Suzaku}/XIS.

\subsection{1FGL\,J2017.3+0603}
\label{subsec:J2017}

\begin{figure}[m]
\begin{center}
\includegraphics[width=150mm]{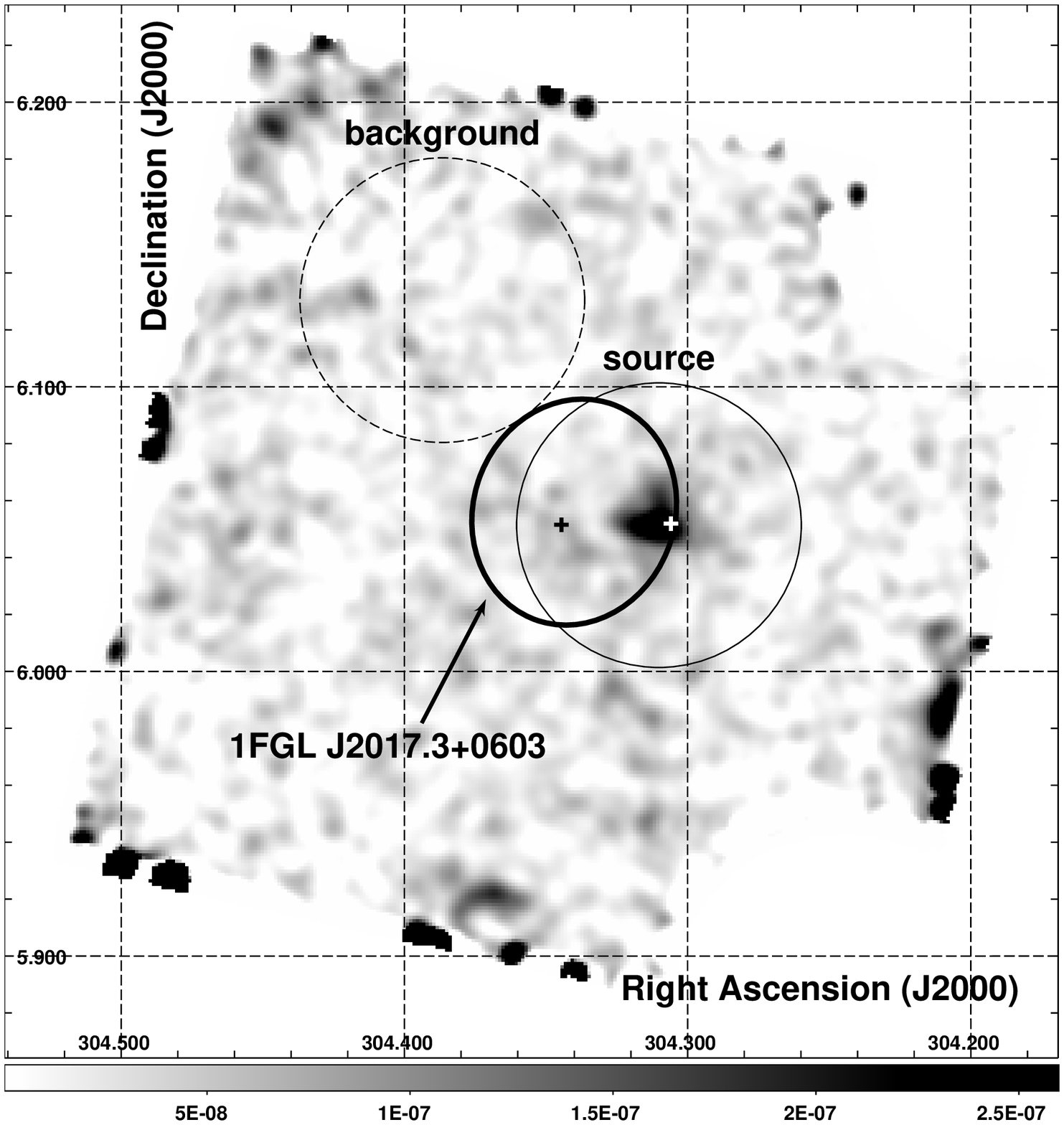} 
\end{center}
\caption{{\it Suzaku}/XIS FI (XIS0+3) image 
 of the 1FGL\,J2017.3$+$0603 region in the $0.4-10$\,keV photon energy
 range.
 The image shows the relative excess of smoothed photon counts
 (arbitrary units indicated in the bottom bar) and is displayed with
 linear scaling.
 The regions enclosed by solid and dashed circles are source and
 background regions, respectively. Thick solid circle denotes $95\%$
 position error of 1FGL\,J2017.3$+$0603. One X-ray point source was
 found within this error circle. White cross mark denotes the position
 of the blazar CLASS\,J2017$+$0603. Black cross mark denotes the
 position of the radio MSP PSR\,J2017$+$0603 \citep{cognard10}.} 
\label{fig:imgJ2017}
\end{figure}

\begin{figure}[m]
\begin{center}
\includegraphics[width=90mm,angle=-90]{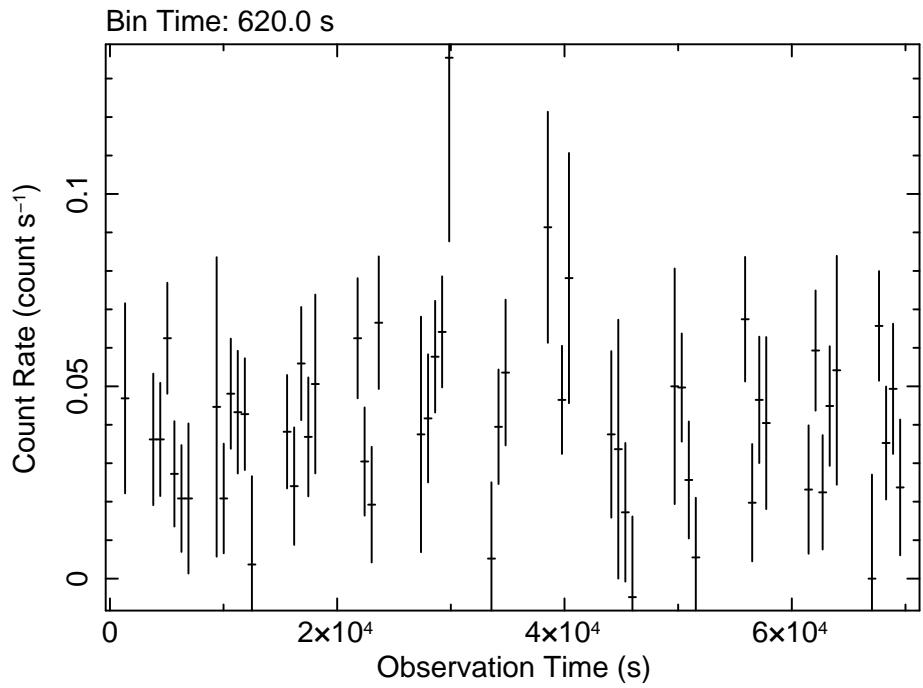} 
\end{center}
\caption{{\it Suzaku}/XIS light curve of an X-ray point source within
 the error circle of 1FGL\,J2017.3$+$0603 with the applied $620$\,s time
 binning. The zero point of time is MJD 55131.4285 (TDB).
} 
\label{fig:lcJ2017}
\end{figure}

\begin{figure}
\begin{center}
\includegraphics[width=90mm,angle=-90]{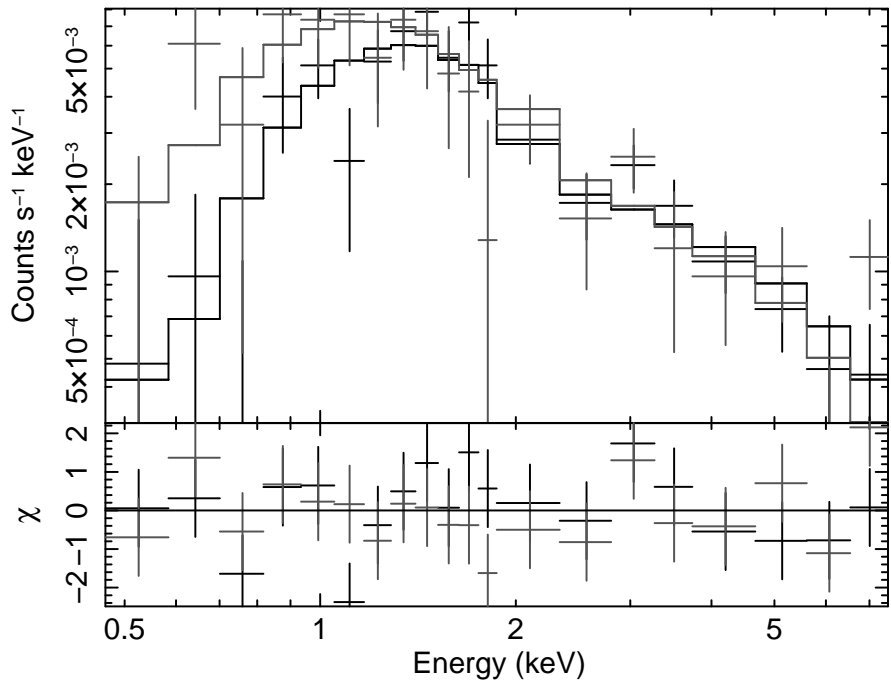} 
\end{center} 
\caption{{\it Suzaku}/XIS spectrum of the potential X-ray counterpart of
 1FGL\,J2017.3$+$0603 with the best fit power-law model. FI data are
 shown black, and BI data in gray.} 
\label{fig:specJ2017}
\end{figure}

A single prominent X-ray point source was found at the edge of
the 1FGL error circle of the unidentified $\gamma$-ray source
1FGL\,J2017.3$+$0603. The X-ray source is located at 
(RA, Dec)\,= (304.\deg310(1), 6.\deg052(1)), as shown in 
Figure~\ref{fig:imgJ2017}. For the further analysis, we set
the extraction region to encircle this bright source with a radius of
$3'$. The corresponding light curve of the newly discovered X-ray source
is show in Figure~\ref{fig:lcJ2017} with $620$\,s binning. The light
curve is consistent (at the level of $>99\%$) with a constant X-ray flux
within the errors ($\chi^2/$d.o.f.\,$=26.4/56$) and the average count
rate $4.07 \times 10^{-2}$\,ct\,s$^{-1}$. Figure~\ref{fig:specJ2017}
shows the X-ray spectrum of the analyzed source. A power-law model
(photon index $\Gamma \simeq 1.6$) with the Galactic absorption provided the
best fit to the data, and the obtained best fit parameters are given in 
Table~\ref{tab:specfitJ2017}. 

The X-ray point source found at the edge of the 1FGL error circle is
positionally coincident (offset by $15''$, which is much less the 
{\it Suzaku}/XIS position accuracy of $\sim 1'$) with the $z=1.743$ 
FSRQ CLASS\,J2017$+$0603 \citep{CLASS}. This blazar has been listed in
the first {\it Fermi}/LAT AGN Catalog \citep{1LAC} as being possibly
associated with 1FGL\,J2017.3+0603, even though the probability for such 
an association was not quantified. We denote its position in
Figure~\ref{fig:imgJ2017} with a white cross. More recently, radio 
and $\gamma$-ray pulsations with the pulse period of $2.9$\,ms have been 
discovered using the Nancay radio telescope well within the {\it Fermi}/LAT 
error 
circle of 1FGL\,J2017.3$+$0603 \citep{cognard10}, pointing instead
to a pulsar (rather than blazar) association of this source.
In Figure~\ref{fig:imgJ2017} we mark the position of the MSP
PSR\,J2017$+$0603 with a black cross. As shown, no X-ray 
counterpart of the pulsar has been detected by {\it Suzaku}/XIS. In
order to determine the corresponding X-ray flux upper limit, we set an
additional source region within $1'$ radius around the position of the
radio pulsar, and assumed a power-law emission spectrum with photon
index $\Gamma = 2$. The resulting 90\% confidence X-ray upper limit is
$S_{2-8\,{\rm keV}} < 2.61 \times 10^{-14}$\,erg\,cm$^{-2}$\,s$^{-1}$.

\section{Discussion}
\label{sec:discussion}

\subsection{The Observed Sample}

\begin{figure}[m]
\begin{center}
\includegraphics[width=150mm]{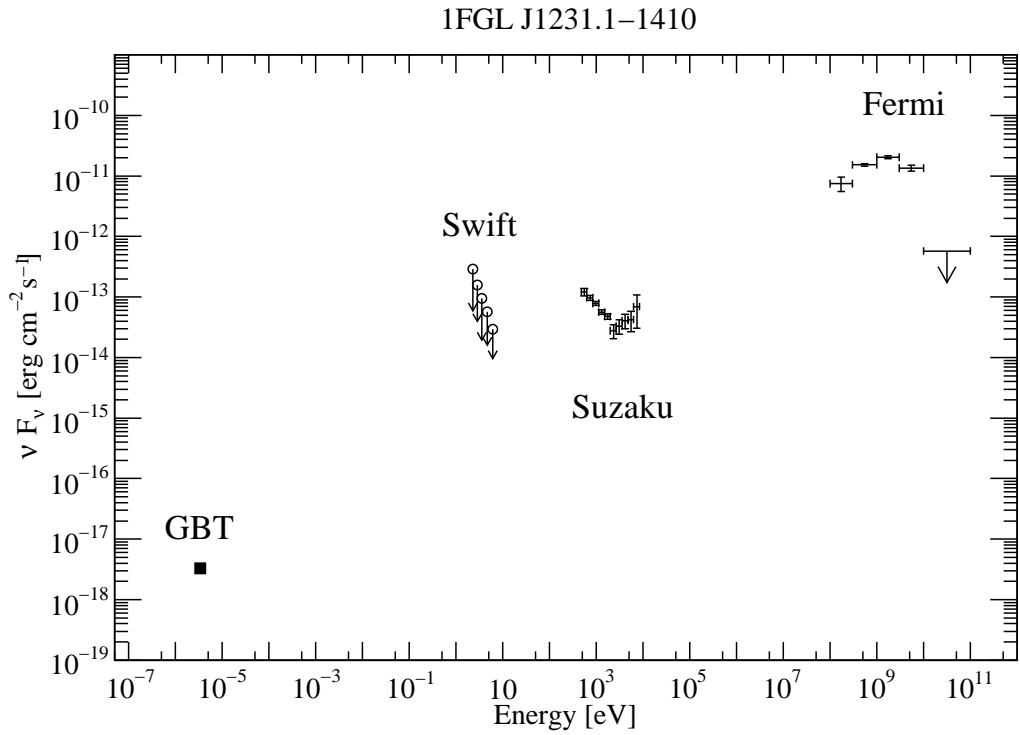} 
\end{center}
\caption{Broad-band spectrum of
 1FGL\,J1231.1$-$1410/PSR\,J1231$-$1411. The X-ray data points represent
 the weighted mean of {\it Suzaku}/XIS FI and BI data. The $\gamma$-ray
 data points are taken from the 1FGL catalog \citep{1FGL}.
 The radio data point is derived from the MSP PSR\,J1231$-$1411 observed
 with Green Bank Telescope by \citet{ransom10}.
 The optical/UV upper limits were derived from the {\it Swift}/UVOT
 observation 
 }
\label{fig:specJ1231-SED}
\end{figure}

Within the error circle of the unidentified $\gamma$-ray object
1FGL\,J1231.4--1410, only one X-ray source was found, which is
positionally consistent with the radio/$\gamma$-ray MSP PSR~J1231--1411
\citep[][see Figure~\ref{fig:imgJ1231}]{ransom10}.
The broad band spectrum of 1FGL\,J1231.1--1410/PSR\,J1231--1411,
including our {\it Suzaku}/XIS data and the derived UVOT optical/UV
upper limits from {\it Swift}, are shown 
in Figure~\ref{fig:specJ1231-SED}. We note that the SED is
reminiscent of that of the Geminga pulsar \citep{thompson99},
or 3EG\,J1835+5918 \citep{halpern02}. Hence the consistent picture emerges, 
in which the $kT\simeq0.16$\,keV blackbody component of the newly 
discovered X-ray counterpart of 1FGL\,J1231.1--1410 originates as 
thermal emission from the surface of a rotating magnetized neutron 
star, a pulsar, while the $\gamma$-ray emission detected by {\it Fermi}/LAT 
may be accounted by the emission of ultra-relativistic electrons 
present within the pulsar magnetosphere. The non-thermal X-ray component 
is then likely to be produced within the magnetosphere of PSR\,J1231--1411 
as well, even though one may also expect some contribution from the outer 
regions (pulsar wind nebulae) to the detected hard X-ray continuum. 

Assuming that PSR\,J1231--1411 is a typical MSP with the
spin period $P = 3.68$\,ms and a spin-down rate 
$\dot{P} = 2.1\times10^{-20}$\,s\,s$^{-1}$ \citep[see][]{ransom10},
one can calculate the corresponding spin-down luminosity as 
$L_{\rm sd}= 4\pi^{2}I\dot{P}\,P^{-3} \simeq 2 \times
10^{34}$\,erg\,s$^{-1}$, and the magnetic field intensity at the 
light cylinder (radius, $R = c P / 2 \pi$) as 
$B_{\rm lc} \simeq 4 \pi^2 (3 I \dot{P} / 2 c^3 P^5)^{1/2} \simeq 5
\times 10^{4}$\,G, where $I = 10^{45}$\,g\,cm$^2$ is the star's 
moment of inertia. Meanwhile, for the claimed distance 
$d \simeq 0.4$\,kpc \citep{ransom10},
the observed $\gamma$-ray luminosity of PSR\,J1231--1411 leads as 
$L_{\gamma} \simeq 2 \times10^{33}$\,erg\,s$^{-1}$, its non-thermal
X-ray luminosity is $L_{\rm X} \simeq 10^{30}$\,erg\,s$^{-1}$, and the
total X-ray luminosity 
$L_{\rm X/tot} \sim 3 \times 10^{30}$\,erg\,s$^{-1}$. These values are
then consistent with the millisecond pulsar scenario --
outer-magnetosphere models in particular -- in a framework of which one
should expect $L_{\gamma} \sim 0.1 \, L_{\rm sd}$ \citep{BSP}
and $L_{\rm X} \sim 10^{-3}\,L_{\rm sd}$ 
\citep{becker97,gaensler06,zhang07},
with relatively large dispersion, however. Interestingly, the
synchrotron X-ray luminosity produced close to the light cylinder within
the expected magnetic field $B_{\rm lc}$ and a fraction (say, $10\%$) of
the volume $V \sim R^3$, would then be close to the observed non-thermal
X-ray luminosity assuming rough energy equipartition between
ultra-relativistic electrons and the magnetic field. 

\begin{figure}[m]
\begin{center}
\includegraphics[width=150mm]{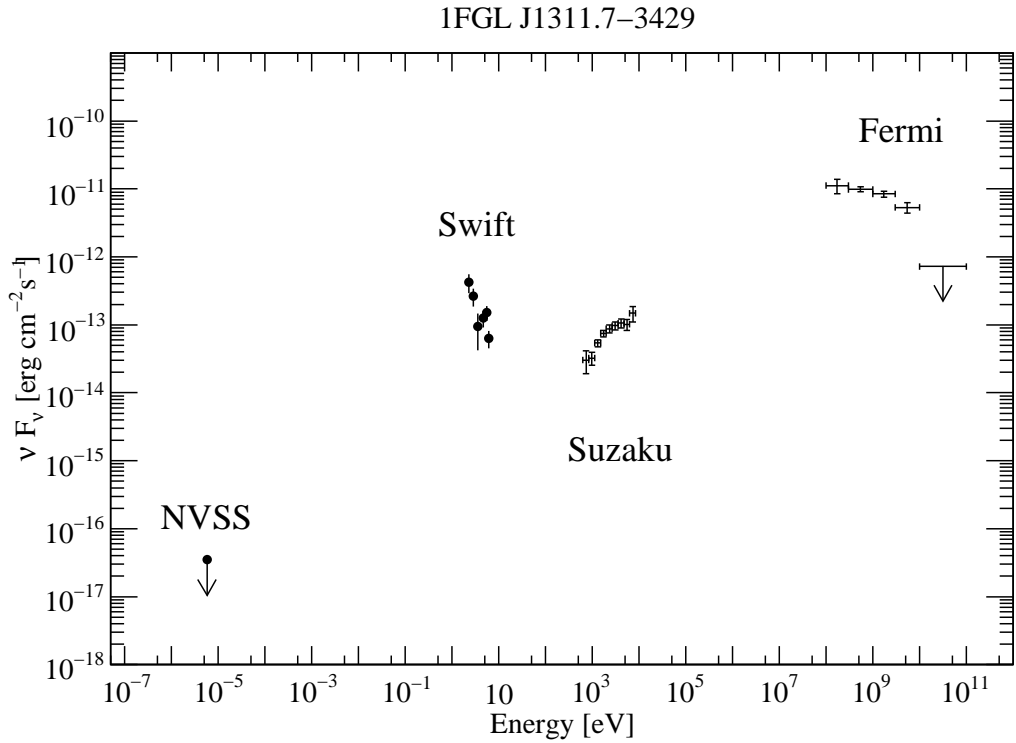} 
\end{center}
\caption{Broad-band spectrum of 1FGL\,J1311.7$-$3429. The X-ray data
 points represent the weighted mean of {\it Suzaku}/XIS FI and BI data
 for src\,A. The $\gamma$-ray data points are taken from the 1FGL
 catalog \citep{1FGL}.
 The radio upper limit is taken from the NVSS catalog \citep{NVSS}.
 The optical/UV data points show the {\it Swift}/UVOT data.}
\label{fig:specJ1311-SED}
\end{figure}

In the case of 1FGL\,J1311.7--3429, two potential X-ray counterparts
have been discovered in our {\it Suzaku} observations. The
association of this {\it Fermi} object with the northern source src\,A
is more likely, since the southern X-ray spot src\,B is located
only marginally within the $95\%$ {\it Fermi}/LAT error circle of 
the $\gamma$-ray emitter.
Yet the classification of 1FGL\,J1311.7--3429/src\,A, 
for which the broad-band spectrum (including radio and optical upper 
limits) is shown in Figure~\ref{fig:specJ1311-SED}, remains vague. 
Currently, no radio or $\gamma$-ray pulsations have been found
at the position of 1FGL\,J1311.7--3429, and this favors an extragalactic 
origin of the detected high-energy emission. And indeed, the flat 
X-ray continuum ($\Gamma\simeq1.4$) and the $\gamma$-ray--to--X-ray energy 
flux ratio $\gtrsim 100$ (with $S_{0.1-10\,{\rm GeV}} \simeq 6.4 \times 10^{-11}$
erg\,cm$^{-2}$\,s$^{-1}$ as given in the 1FGL catalog) would be
consistent with the characteristics of luminous blazars of the FSRQ type
\citep[e.g.,][]{sikora09}.
On the other hand, however, the radio upper limit indicating the GHz
energy flux $\simeq 10^{-5}$ times smaller than the GeV energy flux,
invalidates the blazar nature of 1FGL\,J1311.7--3429. That is because
all active galaxies established till now as $\gamma$-ray emitters are
characterized by relatively strong, Doppler-boosted radio emission. 
In particular, radio energy fluxes of bona fide blazars included in 0FGL
are, for a given {\it Fermi}/LAT photon flux of 
$\sim 10^{-7}$\,photons\,cm$^{-2}$\,s$^{-1}$, at least an order of
magnitude higher than the energy flux implied by the NVSS upper limits
for src\,A \citep[see, e.g.,][]{kovalev09}.
In addition, a very prominent $10$\,ks-long X-ray flare detected from
src\,A, together with the steady GeV flux of 1FGL\,J1311.7--3429, would
not match easily a typical behavior of FSRQs: this class of blazars is
known for displaying dramatic variability at GeV photon energies, but
only modest variations in the X-ray band. Therefore, the nature of the
analyzed {\it Fermi} source and its newly discovered {\it Suzaku}
counterpart remains an open question.

\begin{figure}[m]
\begin{center}
\includegraphics[width=150mm]{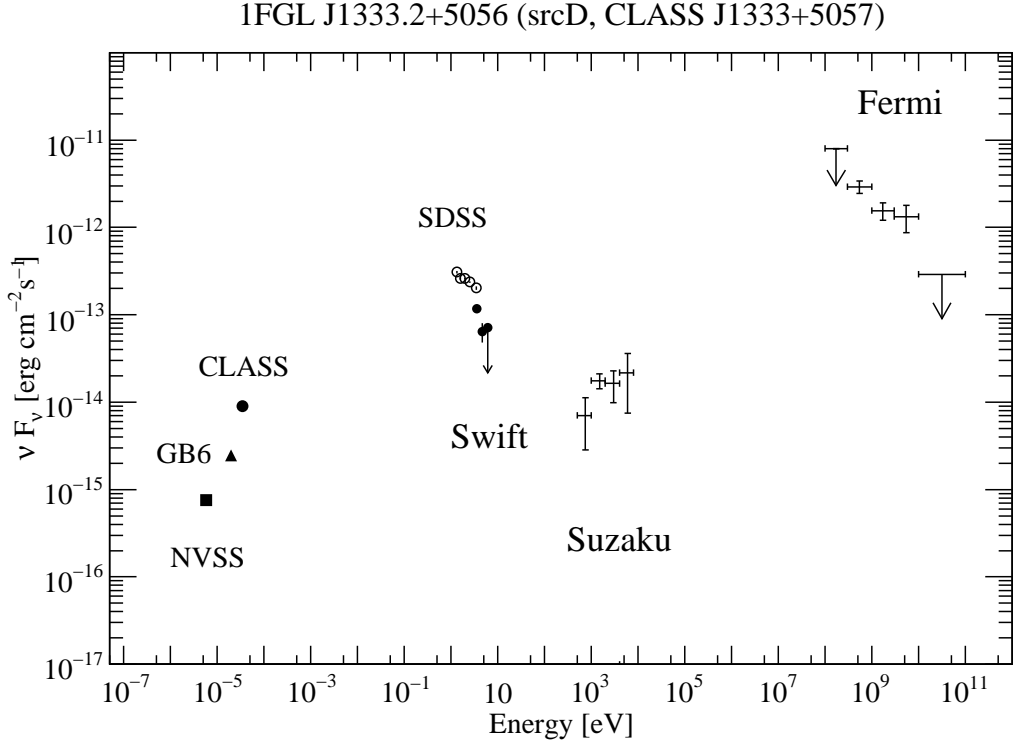} 
\end{center}
\caption{Broad-band spectrum of
 1FGL\,J1333.2+5056/CLASS\,J1333$+$5057. The X-ray data points 
 represent the weighted mean of {\it Suzaku}/XIS FI and BI data for
 src\,D which coincides with the CLASS source. The $\gamma$-ray
 data points are taken from the 1FGL catalog \citep{1FGL}.
 The radio data points, representing blazar CLASS\,J1333$+$5057, are
 taken from the CLASS catalog \citep[filled circle;][]{CLASS},
 NVSS catalog \citep[filled square;][]{NVSS}
 and GB6 catalog \citep[filled triangle;][]{GB6}.
 Optical data point (open circle) was derived from SDSS
 J133353.78$+$505735.9 
 \citep[SDSS Data Release 6;][]{sdssDR6},
 optical/UV data points and upper limit (filled circle) is
 the {\it Swift}/UVOT data.} 
\label{fig:specJ1333-SED}
\end{figure}

Within the error circle of 1FGL\,J1333.2$+$5056, our {\it Suzaku}/XIS
observations revealed the presence of several weak X-ray flux maxima with
possibly diverse spectral properties (as indicated by the spectral
analysis hampered by the limited photon statistics). One of the detected
X-ray sources (src\,D) coincides with the high-redshift blazar
CLASS\,J1333+5056 ($z=1.362$). The broad-band spectral energy
distribution of 1FGL\,J1333.2$+$5056/CLASS\,J1333+5056/src\,D is
presented in Figure~\ref{fig:specJ1333-SED}, including the LAT
$\gamma$-ray, {\it Suzaku} X-ray, archival radio, and newly analyzed
{\it Swift}/UVOT data for the blazar. The constructed SED reveals two 
distinct radiative components, consisting of a
low-energy synchrotron bump and an (energetically dominant)
high-energy inverse-Compton continuum, reminiscent of typical
broad-band spectra for blazars of the FSRQ type \citep{ghisellini98}.
Note that the X-ray--to--$\gamma$-ray flux ratio $\simeq 10^3$ implied by
Figure~\ref{fig:specJ1333-SED}, as well as the relatively large radio flux,
would be both in agreement with the blazar identification of
1FGL\,J1333.2$+$5056. In addition, we note that the discussed 
{\it Fermi} object is the most variable in $\gamma$-rays out of all four
{\it Fermi} targets studied in this paper, 
with the variability index of 38 \citep[which indicates a $<1\%$ 
probability of a steady flux; see][]{1FGL}. 
The additional support for the blazar association is offered by the
fact that the $\gamma$-ray continuum of 1FGL\,J1333.2$+$5056 is the steepest 
among the four {\it Fermi} 
objects observed by us, with the photon index 
$\simeq 2.5 \pm 0.1$, which is compatible with the mean 
$\gamma$-ray photon index of the FSRQ population reported in the 1FGL,
namely $2.47\pm0.19$ \citep{abdo10-FSRQ-BLLac}.

\begin{figure}[m]
\begin{center}
\includegraphics[width=150mm]{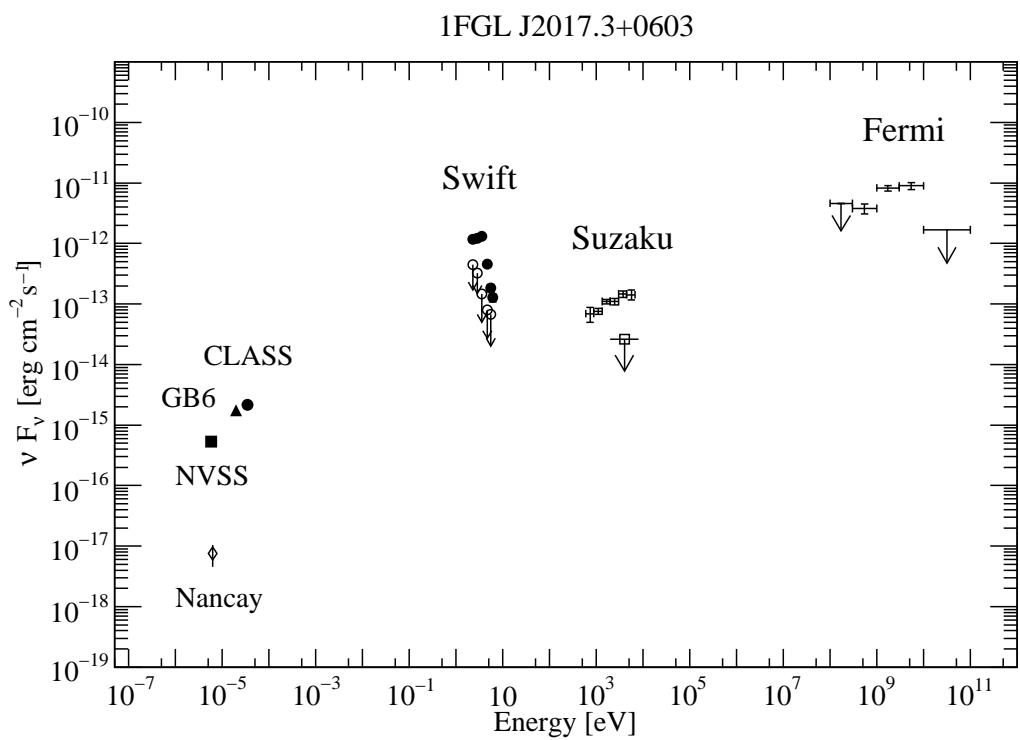} 
\end{center}
\caption{Broad-band spectrum of 1FGL\,J2017.3$+$0603. The X-ray data
 points represent the weighted mean of {\it Suzaku}/XIS FI and BI data
 for active galaxy CLASS\,J2017$+$0603. The X-ray upper limit (open
 square) is derived from the location of the MSP PSR\,J2017$+$0603. The
 $\gamma$-ray data points are taken from the 1FGL catalog \citep{1FGL}. 
 The radio data points, representing CLASS\,J2017$+$0603, are taken from
 the CLASS catalog \citep[filled circle;][]{CLASS},
 NVSS catalog \citep[filled square;][]{NVSS}
 and GB6 catalog \citep[filled triangle;][]{GB6}.
 The open diamond shaped point in radio shows the MSP
 PSR\,J2017+0603 observed with Nancay Radio Telescope \citep{cognard10}
 and also the optical/UV upper limits (open circle) show the MSP
 observed with {\it Swift}/UVOT \citep{cognard10}. The optical/UV data
 points (corresponding filled circle) show the blazar CLASS\,J2017+0603
 observed with {\it Swift}/UVOT.}
\label{fig:specJ2017-SED}
\end{figure}

Finally, in the case of 1FGL\,J2017.3+0603, 
the MSP PSR\,J2017$+$0603 was newly discovered by the Nancay Radio 
Telescope \citep{cognard10}, and the association between the radio and 
$\gamma$-ray sources was confirmed by the pulse detection with the same 
period in the LAT data.
Interestingly, in our {\it Suzaku}/XIS exposure we have only detected 
the high-redshift blazar ($z=1.743$) CLASS\,J2017$+$0603, 
but not the pulsar. The same is true
for the {\it Swift}/UVOT observation \citep{cognard10}, which resulted 
in analogous flux and upper limit measurements in the optical for the 
blazar and pulsar, respectively. The constructed radio to X-ray
SEDs for the 
pulsar and blazar 
systems are shown in Figure~\ref{fig:specJ2017-SED} together with the LAT spectrum. 
Regarding 
the pulsar, \citet{cognard10} discovered that PSR\,J2017+0603 
is located at a distance $d \simeq 1.6$\,kpc, and as such is 
characterized by the spin-down luminosity $L_{\rm sd} \sim 1.34 \times 
10^{34}$\,erg\,s$^{-1}$. The X-ray ($2-8$\,keV) luminosity derived from 
the {\it Suzaku}/XIS upper limit for this pulsar, 
$L_{\rm X} < 8.0 \times 10^{30}$\,erg\,s$^{-1}$, is then consistent with
the expected ``pulsar-like'' luminosity ratio 
$L_{\rm X}/L_{\rm sd} \sim 10^{-3}$.
The 
overall curved
$\gamma$-ray spectrum of 1FGL\,J2017.3+0603,
characterized by the 
small 
photon index $\simeq 1.88\pm0.05$, supports 
the pulsar association. On the other hand, the relatively large radio 
flux of CLASS\,J2017$+$0603, together with the X-ray--to--$\gamma$-ray flux
ratio $\simeq 300$ for the 1FGL\,J2017.3+0603/CLASS\,J2017$+$0603 system,
are in some level of agreement with the blazar interpretation. 
The $\gamma$-ray photon index of 1FGL\,J2017.3$+$0603 is however rather 
flat for a FSRQ and represents a $\sim3\sigma$ deviation from the distribution 
observed for FSRQs \citep[mean$=2.47$, $\sigma=0.19$; see][]{abdo10-FSRQ-BLLac} 
thus making the association with the FSRQ less likely.
Although the detected pulsations in radio and $\gamma$-rays is key to 
the identification of the $\gamma$-ray source with a pulsar, there may be 
some contaminating flux from the blazar. Indeed, the chance probability of 
finding a CLASS-like background blazar in the Fermi error circle of this source 
is $\sim0.003\%$. Considering over 1400 sources in the 1FGL catalog, such 
`mixed' cases could be expected.

\subsection{Implications}

What class of astrophysical objects can be in general associated with
the unidentified high Galactic latitude $\gamma$-ray sources? It was
noted, for example, that compact and relatively nearby molecular clouds
exist at $|b|> 10^{\circ}$, and these should emit $\gamma$-rays at
least at some level. \citet{torres05}
argued, however, that the expected GeV emission of such clouds is too
low to account for the observed fluxes of unidentified EGRET sources,
and the same applies to the bright unidentified {\it Fermi}/LAT
objects. Another classes of possible counterparts proposed were
radio-quiet pulsars and isolated neutron stars \citep[e.g.,][]{yadigaroglu95},
and this idea has indeed been validated by the subsequent multi-frequency
studies, as discussion in $\S$~\ref{sec:intro}. We note in this
context that the Galactic origin of high-latitude $\gamma$-ray emitters
is especially probable for the objects located at 
$10\deg \leq |b| \leq 30^{\circ}$ within the Gould Belt ($\sim 0.3$\,kpc
from the Earth), which constitutes an aggregation of massive late-type
stars, molecular clouds, and supernova remnants \citep{grenier00}.

A probably more challenging population of $\gamma$-ray emitters is
represented by the isotropic component of the unidentified EGRET
objects, consisting of about 60 sources 
\citep[about one third of which with the Galactic latitudes 
$|b|>45\deg$, including several non/weakly-variable during the EGRET
observations;][]{ozel96,gehrels00}. For those sources, \citet{totani00}
have for example suggested associations with large-scale shocks
produced during the structure formation in the intergalactic medium 
\citep[see also][]{waxman00}.
Totani \& Kitayama explored the connection between steady GeV objects
located off the Galactic plane, and labeled in the 3EG catalog as
``possibly extended,'' with dynamically forming clusters of galaxies
\citep[and not single virialized cluster systems; see][]{kawasaki02}.
However, the non-variable nature of the $\gamma$-ray emission of several
of the considered objects was questioned \citep[see][and references 
therein]{reimer03}, and the high efficiency of the particle acceleration 
at the structure formation shocks required by the model was also noted 
\citep[e.g.,][]{keshet03}.

Radio galaxies are prime candidates for the unidentified high Galactic
latitude EGRET sources, especially since the only confirmed non-blazar
AGN detected previously at GeV photon energies was the nearby radio
galaxy Centaurus\,A \citep{steinle98,sreekumar99}.
Yet no other radio galaxy has been firmly detected by EGRET at the
significance level high enough ($\geq 4\sigma$) to be included in the
3rd EGRET catalog \citep{3EG}. Moreover, \citet{cillis04},
who applied a stacking analysis of the EGRET data for a sample of the
brightest and/or the closest radio and Seyfert galaxies, showed that `no
detection significance greater than $2\sigma$ has been found for any
subclass, sorting parameter, or number of objects co-added.'
Nevertheless, \citet{mukherjee02} argued that the most likely counterpart 
to the unidentified EGRET source 3EG\,J1621+8203 is the bright radio galaxy 
NGC\,6251. A marginal detection of 3C111 with EGRET has also been reported
\citep{hartman08}. We also note that \citet{combi03} reported the 
discovery of a new radio galaxy within the location error circle of 
the unidentified $\gamma$-ray source 3EG\,J1735--1500. The
identification of 3EG\,J1735--1500 was however controversial, due to
the presence of an another likely (blazar-type) candidate within the
EGRET error contours \citep{sowards04}.
The most recent analysis based on the 15 months of {\it Fermi}/LAT data
resulted in the detection of 11 non-blazar-type AGN (all radio galaxies),
including the aforementioned cases of NGC\,6251 and 3C111 
\citep{abdo10-AGNpress}. 
The idea that some fraction of unidentified $\gamma$-ray emitters may be 
associated with faint radio galaxies is therefore validated, although this
should rather apply to only dimmer {\it Fermi} objects, and not to the 
population of exceptionally bright $\gamma$-ray sources detected already 
by EGRET.

The {\it Suzaku}/XIS studies of four bright {\it Fermi}/LAT objects
reported here provide an important contribution to the debate regarding
the nature of unidentified $\gamma$-ray emitters located at high
Galactic latitudes. In particular, our observations 
support the idea
that a significant fraction of such objects may be associated with old
($\gtrsim$\,Gyr) MSPs present within the Galactic halo and the Earth's
neighborhood (such as 1FGL\,J1231.1--1410 and 1FGL\,J2017.3+0603). 
Yet not all of the unidentified {\it Fermi} objects are related to the
pulsar phenomenon. Instead, some of those may be hosted by active
galaxies, most likely by the luminous and high-redshift blazars of the
FSRQ type (1FGL\,J1333.2+5056 is as good blazar candidate, for 
example). However, there still remain unidentified sources,
(e.g., 1FGL\,J1311.7--3429), for which neither blazar nor pulsar scenarios
seem to apply. For these, ultra-deep multi-wavelength studies are 
probably needed to unravel their physical nature.

\section{Summary}
\label{sec:summary}

In this paper we reported on the results of deep X-ray follow-up observations
of four unidentified $\gamma$-ray sources detected by the 
{\it Fermi}/LAT instrument at high Galactic latitudes 
($|b| > 10^{\circ}$) using the X-ray Imaging Spectrometers onboard 
{\it Suzaku} satellite. All of the studied objects have been detected at
high significance ($> 10 \sigma$) during the first 3-months of the 
{\it Fermi}/LAT operation.
For some of them, possible associations with pulsars and blazar
have been recently discussed, and our observations provide an important
contribution to this debate. In particular, an X-ray point source was
newly found within $95\%$ error circle of 1FGL\,J1231.1--1410. The
X-ray spectrum of the discovered {\it Suzaku} counterpart of
1FGL\,J1231.1--1410 is well fitted by a blackbody emission with a
temperature of $kT \simeq 0.16$\,keV plus an additional power-law
component with a differential photon index $\Gamma \simeq 1.8$. This
supports the recently claimed identification of this source with a MSP
PSR\,J1231--1411. For the remaining three {\it Fermi} objects, the
performed X-ray observations are less conclusive. In the case of
1FGL\,J1311.7--3429, two possibly associated X-ray point sources were 
newly found. Even though the $0.4-10$\,keV spectral and
variability properties for those could be robustly accessed, the
physical nature of the X-ray emitters and their relations with the
$\gamma$-ray source remain unidentified. Similarly, we found
several weak X-ray sources in the field of 1FGL\,J1333.2+5056, one
coinciding with the high-redshift blazar CLASS\,J1333+5057. We argue
that the available data are consistent with the physical association
between these two objects, even though we were not able to identify
robustly the {\it Suzaku} counterpart of $\gamma$-ray emitter due to a
large positional uncertainty of 1FGL\,J1333.2+5056. Finally, we 
found an X-ray point source in the vicinity of 1FGL\,J2017.3+0603. This
{\it Fermi} object was recently suggested to be associated with a newly
discovered MSP PSR\,J2017+0603 because of the detection of radio 
and $\gamma$-ray pulsations. However, we did not detect the X-ray 
counterpart of the pulsar, but instead of the high-redshift blazar 
CLASS\,J2017+0603 located within the error circle of 1FGL\,J2017.3+0603. 
Still, the resulting upper limits for the X-ray emission do not 
invalidate the pulsar association.

\acknowledgments
{\L}.S. is grateful for the support from Polish MNiSW through the grant
N-N203-380336.

\clearpage

\begin{table}[m]
\small
\caption{EGRET and {\it Fermi}/LAT entries for the analyzed objects}
\label{tab:3EG-1FGL}
\begin{center}
\begin{tabular}{ccccccc}
\hline \hline
Name & RA & DEC & $l$ & $b$ & $F_{\rm 0.1-20\,GeV}$ & $r_{\rm 95\%}$ \\
 & [deg] & [deg] & [deg] & [deg] & [$10^{-8}$\,ph\,cm$^{-2}$\,s$^{-1}$] & [deg] \\
\hline \hline
  1FGL\,J1231.1$-$1410$^{\dag}$ & 187.80 & $-$14.17 & 295.53 & $+$48.41 & 14.9$\pm$0.7 & 0.03 \\ 
  3EG\,J1234$-$1318 & 188.19 & $-$16.30 & 296.43 & $+$49.34 & 21.6$\pm$5.3 & 0.76 \\ 
  (EGR\,J1231$-$1412) & & & & & & \\
\hline
  1FGL\,J1311.7$-$3429$^{\S}$ & 197.95 & $-$34.49 & 307.69 & $+$28.19 & 11.7$\pm$1.1 & 0.04 \\ 
  3EG\,J1314$-$3431 & 198.51 & $-$34.52 & 308.21 & $+$28.12 & 18.7$\pm$3.1 & 0.56 \\ 
  (EGR\,J1314$-$3417) & & & & & & \\
\hline
  1FGL\,J1333.2$+$5056$^{\S}$ & 203.30 & $+$50.94 & 107.32 & $+$64.90 & 4.5$\pm$1.0 & 0.15 \\ 
  3EG\,J1337$+$5029 & 204.39 & $+$50.49 & 105.40 & $+$65.04 & 9.2$\pm$2.6 & 0.72 \\ 
  (EGR\,J1338$+$5102) & & & & & & \\
\hline
  1FGL\,J2017.3$+$0603$^{\ddag}$ & 304.34 & $+$6.05 & 48.62 & $-$16.02 & 4.5$\pm$0.5 & 0.04 \\ 
\hline
\end{tabular}
\end{center}
 {\small $^{\dag}$ Data consistent with no variability between EGRET and {\it Fermi}/LAT observations.} \\ 
 {\small $^{\S}$ The $\gamma$-ray fluxes determined by EGRET and {\it Fermi}/LAT marginally consistent within $2\sigma$ level.} \\ 
 {\small $^{\ddag}$ Data consistent with no variability between EGRET and {\it Fermi}/LAT observations because of the EGRET detection limit $\simeq 6 \times10^{-8}$\,ph\,cm$^{-2}$\,s$^{-1}$.}
\end{table}

\clearpage

\begin{table}[m]
\small
\caption{{\it Suzaku}/XIS Observation Log} 
\label{tab:exposure}
\begin{center}
\begin{tabular}{cccccc}
\hline \hline
\multicolumn{1}{c}{Name} & \multicolumn{1}{c}{OBS ID} &
 \multicolumn{2}{c}{Pointing Center$^{*}$} & Observation start & Effective exposure \\ 
  &   & \multicolumn{1}{c}{RA [deg]} & \multicolumn{1}{c}{DEC [deg]} & (UT) & [ksec] \\ 
\hline \hline
  1FGL\,J1231.1$-$1410 & 804017010$^{\dag}$ & $187.8001$ & $-14.1665$ & 2009/07/08 22:53:48 & 23.8 \\  
	& 804017020$^{\dag}$ & $187.7993$ & $-14.1672$ & 2009/07/28 05:21:37 & 44.8 \\  
  1FGL\,J1311.7$-$3429 & 804018010 & $197.9603$ & $-34.4918$ & 2009/08/04 04:56:35 & 33.0 \\  
  1FGL\,J1333.2$+$5056 & 804019010 & $203.2955$ & $51.0170$ & 2009/06/01 10:13:15 & 39.1 \\  
  1FGL\,J2017.3$+$0603 & 804020010 & $304.3461$ & $6.0496$ & 2009/10/27 10:14:45 & 26.7 \\
\hline
\end{tabular}
\end{center}
{\small $^{*}$ The pointing centers were the positions given in 0FGL catalog \citep{0FGL}.} \\ 
{\small $^{\dag}$ The requested continuous $80$\,ks {\it Suzaku} exposure was interrupted by Target of Opportunity (ToO) observation of GRB\,090708. For this reason the observation was divided into $30$\,ks and $50$\,ks segments before and after the ToO observation.} 
\end{table}

\clearpage

\begin{table}[m]
\caption{Source detection results of {\it Suzaku} observation} 
\label{tab:srcdetect}
\begin{center}
\begin{tabular}{cccccc}
\hline \hline
Name & & \multicolumn{2}{c}{Position} & Detection Significance & $r_{95\%}$ \\
     & & RA [deg] & DEC [deg] & $\sigma$ & [arcsec] \\
\hline \hline
1FGL\,J1231.4--1410 & ---    & $187.790$ & $-14.192$ & $13.67$ &  $7.44$ \\
1FGL\,J1311.7--3429 & src\,A & $197.939$ & $-34.508$ & $15.52$ & $17.44$ \\
                    & src\,B & $197.942$ & $-34.534$ & $12.89$ & $12.27$ \\
1FGL\,J1333.2+5056  & src\,A & $203.252$ &  $50.983$ &  $8.53$ & $23.34$ \\
                    & src\,B & $203.161$ &  $51.032$ &  $7.27$ & $19.97$ \\
                    & src\,C & $203.276$ &  $51.014$ &  $7.47$ & $20.75$ \\
                    & src\,D & $203.479$ &  $50.967$ &  $4.50$ & $38.41$ \\
                    & src\,E & $203.381$ &  $50.892$ &  $4.91$ & $26.51$ \\
1FGL\,J2017.3+0603  & ---    & $304.310$ &   $6.052$ & $14.44$ &  $4.73$ \\
\hline
\end{tabular}
\end{center}
\end{table}

\clearpage

\begin{table}[m]
\small
\caption{Fitting Parameters for 1FGL\,J1231.1$-$1410 in a framework of blackbody (BB) and power-law (PL) models}
\label{tab:specfitJ1231}
\begin{center}
\begin{tabular}{ccccc}
\hline \hline
          & BB model & BB$+$PL Model \\ 
parameter & value \& error & value \& error \\
\hline \hline
  $N_H$ [$10^{22}$\,cm$^{-2}$] & $0.0$ (fixed) & $0.0$ (fixed) \\
  $kT$ [keV] & 0.228 $\pm$ 0.008 & $0.163~^{+0.024}_{-0.026}$ \\
  norm. (BB) & $(1.42 \pm 0.14) \times 10^{-6}$ & $(1.20~^{+0.31}_{-0.37}) \times 10^{-6}$ \\
  $\Gamma$ & --- & $1.79$ $^{+0.40}_{-0.17}$ \\
  norm. (PL) & --- & $(1.94~^{+1.14}_{-0.84}) \times 10^{-5}$ \\ 
\hline
 $\chi^2$ & $128.1$ & $55.46$ \\
 d.o.f.  &  $34$    & $32$ \\
 reduced $\chi^2$ & $3.768$ & $1.733$ \\
\hline
 Flux $(2-8\,{\rm keV})$ & --- & $(5.79~^{+1.62}_{-1.52})\times 10^{-14}$ \\
 $[$erg\,cm$^{-2}$\,s$^{-1}$ $]$ & \multicolumn{2}{c}{ } \\
\hline
\end{tabular}
\end{center}
\end{table} 

\clearpage

\begin{table}[m]
\small
\caption{Blackbody (BB) and power-law (PL) components in the X-ray spectrum of 1FGL\,J1231.1$-$1410}
\label{tab:ftest} 
\begin{center}
\begin{tabular}{ccccccc}
\hline \hline
  & \multicolumn{2}{c}{(i) Standard Background} & \multicolumn{2}{c}{(ii) Ring Background} & \multicolumn{2}{c}{(iii) Lockman Hole Background} \\  
  & BB & BB$+$PL & BB & BB$+$PL & BB & BB$+$PL \\
\hline \hline
  $\chi^2$ & 134.71 & 56.16 & 67.21 & 18.97 & 39.77 & 23.97 \\
  d.o.f.   & 34 & 32 & 30 & 28 & 38    & 36    \\
\hline
  F value     & \multicolumn{2}{c}{22.4} & \multicolumn{2}{c}{35.6} &
 \multicolumn{2}{c}{11.9} \\ 
  Probability & \multicolumn{2}{c}{8.33$\times10^{-7}~\%$} &
	     \multicolumn{2}{c}{2.04$\times10^{-6}~\%$} &
 \multicolumn{2}{c}{1.10$\times10^{-2}~\%$} \\ 
\hline
\end{tabular}
\end{center}
\end{table}

\clearpage

\begin{table}[m]
\small
\caption{Fitting Parameters for 1FGL\,J1311.7$-$3429 for power-law model}
\label{tab:specfitJ1311}
\begin{center}
\begin{tabular}{ccc}
\hline \hline
  & src\,A & src\,B \\
  parameter & value \& error & value \& error \\
\hline \hline
  $N_H$ [$10^{20}$\,cm$^{-2}$]& $4.45$ (fixed) & $4.45$ (fixed) \\ 
  $\Gamma$ & $1.38~^{+0.13}_{-0.13}$ & $1.34~^{+0.16}_{-0.15}$ \\
  norm. & $(2.69~^{+0.38}_{-0.37})\times 10^{-5}$ & $(2.08~^{+0.34}_{-0.33})\times 10^{-5}$ \\ 
\hline
  $\chi^2$ & $42.6$ & $42.1$ \\
  d.o.f.   & $38$   & $38$ \\
  reduced $\chi^2$ & $1.12$ & $1.11$ \\ 
\hline
 Flux ($2-8$\,keV) & $(1.45~^{+0.18}_{-0.18}) \times 10^{-13}$ & $(1.20~^{+0.18}_{-0.17}) \times 10^{-13}$ \\
 $[$ erg\,cm$^{-2}$\,s$^{-1}$ $]$ \\
\hline
\end{tabular}
\end{center}
\end{table}

\clearpage

\begin{table}[m]
\small
\caption{Count rates and constant flux fits for X-ray sources within the error circle of 1FGL\,J1333.2+5056} 
\label{tab:lcfitJ1333}
\begin{center}
\begin{tabular}{cccc}
\hline \hline
Source & Average count rate \& Error & $\chi^2/$d.o.f. & Prob. \\
       & [ $10^{-3}$ ct\,s$^{-1}$] & & [\%]\\
\hline \hline
  src\,A & $5.47 \pm 0.51$ & $14.9/15$ & 46.08 \\ 
  src\,B & $4.40 \pm 0.49$ & $22.4/15$ & 9.73 \\ 
  src\,C & $4.37 \pm 0.48$ & $18.8/15$ & 22.28 \\ 
  src\,D & $2.19 \pm 0.44$ & $20.5/15$ & 15.25 \\ 
  src\,E & $1.70 \pm 0.44$ & $23.3/15$ & 7.86 \\ 
\hline
\end{tabular}
\end{center}
\end{table}

\clearpage

\begin{table}[m]
\small
\caption{Fitting Parameters for 1FGL\,J1333.2+5056 for power-law model}
\label{tab:specfitJ1333}
\begin{center}
\begin{tabular}{cccccc}
\hline \hline
            & src\,A         & src\,B         & src\,C         & src\,D         & src\,E \\ 
  parameter & value \& error & value \& error & value \& error & value \& error & value \& error \\ 
\hline \hline
  $N_H$ [$10^{20}$\,cm$^{-2}$] & $1.09$ (fixed) & $1.09$ (fixed) & $1.09$ (fixed) & $1.09$ (fixed) & $1.09$ (fixed) \\ 
  $\Gamma$ & $2.35~^{+0.35}_{-0.32}$ & $1.48~^{+0.29}_{-0.27}$ & $1.51~^{+0.31}_{-0.29}$ & $1.76~^{+0.60}_{-0.52}$ & $2.50~^{+0.69}_{-0.58}$ \\ 
  norm. [$\times10^{-5}$] & $1.57~^{+0.28}_{-0.28}$ & $1.07~^{+0.27}_{-0.26}$ & $0.84~^{+0.22}_{-0.22}$ & $0.77~^{+0.31}_{-0.30}$ & $1.34~^{+0.36}_{-0.37}$ \\ 
\hline
  $\chi^2$ & $13.0$ & $7.33$ & $18.3$ & $12.7$ & $11.4$ \\ 
  d.o.f.   & $18$   & $18$   & $18$   & $16$   & $12$ \\ 
  reduced $\chi^2$ & $0.720$ & $0.407$ & $1.02$ & $0.796$ & $0.949$ \\ 
\hline
 Flux ($2-8$\,keV) & $2.16~^{+0.88}_{-0.75}$ & $4.98~^{+1.46}_{-1.37}$ & $3.77~^{+1.17}_{-1.11}$ & $2.41~^{+1.55}_{-1.26}$ & $1.52~^{+1.41}_{-0.94}$ \\
 $[\times 10^{-14}$ erg\,cm$^{-2}$\,s$^{-1}$ $]$ \\
\hline
\end{tabular}
\end{center}
\end{table}

\clearpage

\begin{table}[m]
\small
\caption{Fitting Parameters for 1FGL\,J2017.3$+$0603 for power-law model}
\label{tab:specfitJ2017}
\begin{center}
\begin{tabular}{cc}
\hline \hline
  parameter & value \& error \\
\hline \hline
  $N_H$ [$10^{22}$\,cm$^{-2}$] & $0.1309$ (fixed) \\
  $\Gamma$ & $1.59~^{+0.15}_{-0.15}$ \\
  norm. & $(5.03~^{+0.68}_{-0.66})\times10^{-5}$ \\
\hline
  $\chi^2$ & $34.8$ \\
  d.o.f.  & $38$ \\
  reduced $\chi^2$ & $0.916$ \\
\hline
 Flux ($2-8$\,keV) & $(1.99~^{+0.28}_{-0.27}) \times 10^{-13}$ \\
 $[$ erg\,cm$^{-2}$\,s$^{-1}$ $]$ \\
\hline
\end{tabular}
\end{center}
\end{table}

\clearpage

\clearpage

\clearpage

\clearpage

\clearpage

\clearpage

\clearpage

\clearpage

\clearpage

\clearpage

\clearpage

\clearpage

\clearpage

\clearpage

\clearpage

\clearpage

\begin{thebibliography}{}
\bibitem[Abdo et al.(2009a)]{LBAS} 
		Abdo, A.~A., et al. ({\it Fermi}-LAT Collaboration)
		2009a, ApJ, 700, 597
\bibitem[Abdo et al.(2009b)]{PSRJ2021} 
		Abdo, A.~A., et al. ({\it Fermi}-LAT Collaboration)
		2009b, ApJ, 700, 1059
\bibitem[Abdo et al.(2009c)]{0FGL} 
		Abdo, A.~A., et al. ({\it Fermi}-LAT Collaboration)
		2009c, \apjs, 183, 46
\bibitem[Abdo et al.(2009d)]{BSP} 
		Abdo, A.~A., et al. ({\it Fermi}-LAT Collaboration)
		2009d, Science, 325, 840
\bibitem[Abdo et al.(2010a)]{PSRJ1836} 
		Abdo, A.~A., et al. ({\it Fermi}-LAT Collaboration)
		2010a, ApJ, 712, 1209
\bibitem[Abdo et al.(2010b)]{PSRJ0034} 
		Abdo, A.~A., et al. ({\it Fermi}-LAT Collaboration)
		2010b, ApJ, 712, 957
\bibitem[Abdo et al.(2010c)]{1FGL} 
		Abdo, A.~A., et al. ({\it Fermi}-LAT Collaboration)
		2010c, \apjs, 188, 405
\bibitem[Abdo et al.(2010d)]{1LAC} 
		Abdo, A.~A., et al. ({\it Fermi}-LAT Collaboration)
		2010d, ApJ, 715, 429
\bibitem[Abdo et al.(2010e)]{1LGPC} 
		Abdo, A.~A., et al. ({\it Fermi}-LAT Collaboration)
		2010e, \apjs, 187, 460
\bibitem[Abdo et al.(2010f)]{abdo10-FSRQ-BLLac}
                Abdo, A.~A., et al. ({\it Fermi}-LAT Collaboration)
                2010f, ApJ, 720, 435
\bibitem[Abdo et al.(2010g)]{abdo10-AGNpress} 
		Abdo, A.~A., et al. ({\it Fermi}-LAT Collaboration)
		2010f, \apj, 720, 912
\bibitem[Adelman-McCarthy et al.(2008)]{sdssDR6} 
		Adelman-McCarthy, J.~K., et al. 2008, ApJS, 175, 297
\bibitem[Atwood et al.(2009)]{atwood09} 
		Atwood, W.~B., et al. 2009, ApJ, 697, 1071
\bibitem[Becker \& Truemper(1997)]{becker97} 
		Becker, W., \& Truemper, J. 1997, A\&A, 326, 682
\bibitem[Casandjian \& Grenier(2008)]{EGR} 
		Casandjian, J.-M., \& Grenier, I.~A. 2008, A\&A, 489, 849
\bibitem[Cillis et al.(2004)]{cillis04} 
		Cillis, A.~N., Hartman, R.~C., \& Bertsch, D.~L. 2004,
		ApJ, 601, 142
\bibitem[Combi et al.(2003)]{combi03} 
		Combi, J.~A., et al. 2003, ApJ, 588, 731
\bibitem[Cognard et al.(2010)]{cognard10} 
		Cognard, I., et al. 2010, submitted
\bibitem[Condon et al.(1998)]{NVSS} 
		Condon, J.~J., et al. 1998, AJ, 115, 1693
\bibitem[Day et al.(1998)]{day98}
		Day, C., et al. 1998, The ASCA Data Reduction Guide,
		Tech. Rep., (Greenbelt: NASA GSFC), v.2.0
\bibitem[Dickey \& Lockman(1990)]{Colden} 
		Dickey, J.~M., \& Lockman, F.~J. 1990, ARA\&A, 28, 215 
\bibitem[Gaensler \& Slane(2006)]{gaensler06} 
		Gaensler, B.~M., \& Slane, P.~O. 2006, ARA\&A, 44, 17
\bibitem[Gehrels \& Michelson(1999)]{gehrels99} 
		Gehrels, N., \& Michelson, P. 1999, Astropart. Phys.,
		11, 277
\bibitem[Gehrels et al.(2000)]{gehrels00} 
		Gehrels, N., et al. 2000, Nature, 404, 363
\bibitem[Ghisellini et al.(1998)]{ghisellini98} 
		Ghisellini, G., et al. 1998, MNRAS, 301, 451
\bibitem[Gregory et al.(1996)]{GB6} 
		Gregory, P.~C., et al. 1996, ApJS, 103, 427
\bibitem[Grenier et al.(2000)]{grenier00} 
		Grenier, I.~A. 2000, A\&A, 364, 93
\bibitem[Halpern et al.(2001)]{halpern01} 
		Halpern, J.~P., et al. 2001, ApJ, 551, 1016
\bibitem[Halpern et al.(2002)]{halpern02} 
		Halpern, J.~P. et al. 2002, ApJ, 573, L41
\bibitem[Halpern et al.(2008)]{halpern08} 
		Halpern, J.~P. et al. 2008, ApJ, 688, L33
\bibitem[Hartman et al.(1999)]{3EG} 
		Hartman, R.~C., et al. 1999, ApJS, 123, 79
\bibitem[Hartman, Kadler \& Tueller(2008)]{hartman08} 
		Hartman, R. C., Kadler, M., \& Tueller, J. 2008,
		ApJ, 688, 852
\bibitem[Ishisaki et al.(2007)]{ishisaki07} 
		Ishisaki, Y., et al. 2007, PASJ, 59, S113
\bibitem[Kawasaki \& Totani(2002)]{kawasaki02} 
		Kawasaki, W., \& Totani, T. 2002, ApJ, 576, 679
\bibitem[Keshet et al.(2003)]{keshet03} 
		Keshet, U., et al. 2003, ApJ, 585, 128
\bibitem[Kokubun et al.(2007)]{kokubun07} 
		Kokubun, M., et al. 2007, PASJ, 59, S53
\bibitem[Kovalev et al.(2009)]{kovalev09} 
		Kovalev, Y.~Y., et al. 2009, ApJ, 696, 17
\bibitem[Koyama et al.(2007)]{koyama07} 
		Koyama, K., et al. 2007, PASJ, 59, S23
\bibitem[Myers et al.(2003)]{CLASS} 
		Myers, S.~T. et al., 2003, MNRAS, 341, 1
\bibitem[Mirabal et al.(2000)]{mirabal00} 
		Mirabal, N., et al. 2000, ApJ, 541, 180
\bibitem[Mitsuda et al.(2007)]{mitsuda07} 
		Mitsuda, K., et al., 2007, PASJ, 59, 1
\bibitem[Morrison \& McCammon(1983)]{morrison83} 
		Morrison, R., \& McCammon, D. 1983, ApJ, 270, 119
\bibitem[Mukherjee et al.(2000)]{mukherjee00} 
		Mukherjee, R., et al. 2000, ApJ, 542, 740 
\bibitem[Mukherjee et al.(2002)]{mukherjee02} 
		Mukherjee, R., et al. 2002, ApJ, 574, 693  
\bibitem[Mukherjee \& Halpern(2004)]{mukherjee04}
                Mukherjee, R., \& Halpern, J. 2004, Cosmic Gamma-Ray Sources, Eds. K.S.~Cheng \& G.E.~Romero, 304, 311 
\bibitem[Nolan et al.(2003)]{nolan03} 
		Nolan, P.~L., et al. 2003, ApJ, 597, 615
\bibitem[\"Ozel \& Thompson(1996)]{ozel96} 
		\"Ozel, M.~E., \& Thompson, D.~J. 1996, ApJ, 463, 105
\bibitem[Ransom et al.(2010)]{ransom10} 
		Ransom, S.~M., et al., 2010, ApJ, 727, L16
\bibitem[Reimer(2001)]{reimer01book} 
		Reimer, O., 2001, in proc. {\it `The Nature of
		Unidentified Galactic High-Energy Gamma-Ray Sources'},
		eds. A. Carraminana, O. Reimer, \& D.~J. Thompson
		(Kluwer Academic Publishers: Dordrecht), 17 
\bibitem[Reimer et al.(2001)]{reimer01} 
		Reimer, O., et al. 2001, MNRAS, 324, 772
\bibitem[Reimer et al.(2003)]{reimer03} 
		Reimer, O., et al. 2003, ApJ, 588, 155
\bibitem[Shaw et al.(2009)]{shaw09}
		Shaw, M.~S., et al. 2009, ApJ, 704, 477
\bibitem[Sikora et al.(2009)]{sikora09} 
		Sikora, M., et al. 2009, ApJ, 704, 38
\bibitem[Sowards-Emmerd et al.(2003)]{sowards03} 
		Sowards-Emmerd, D., et al. 2003, ApJ, 590, 109
\bibitem[Sowards-Emmerd et al.(2004)]{sowards04} 
		Sowards-Emmerd, D., Romani, R.~W., \& Michelson,
		P.~F. 2004, ApJ, 609, 564
\bibitem[Sreekumar et al.(1999)]{sreekumar99} 
		Sreekumar, P., et al. 1999, APh, 11, 221
\bibitem[Steinle et al.(1998)]{steinle98} 
		Steinle, H., et al., 1998, A\&A, 330, 97
\bibitem[Takahashi et al.(2007)]{takahashi07} 
		Takahashi, T., et al. 2007, PASJ, 59S, 35
\bibitem[Tavani et al.(2009)]{AGILE} 
		Tavani, M., et al. 2009, A\&A, 502, 995
\bibitem[Tawa et al.(2008)]{tawa08} 
		Tawa, N., et al. 2008, PASJ, 60S, 11
\bibitem[Thompson et al.(1999)]{thompson99} 
		Thompson, D.~J., et al. 1999, ApJ, 516, 297
\bibitem[Torres et al.(2005)]{torres05} 
		Torres, D.~F., Dame, T.~M., \& Digel, S.~W. 2005, ApJ, 621, L29 
\bibitem[Totani \& Kitayama(2000)]{totani00} 
		Totani, T., \& Kitayama, T. 2000, ApJ, 545, 572
\bibitem[Waxman \& Loeb(2000)]{waxman00} 
		Waxman, E \& Loeb, E. 2000, ApJ, 545, L11
\bibitem[Yadigaroglu \& Romani(1995)]{yadigaroglu95} 
		Yadigaroglu, I.~A., \& Romani, R.~W. 1995, ApJ, 449, 211
\bibitem[Zhang et al.(2007)]{zhang07} 
		Zhang, L., et al. 2007, ApJ, 666, 1165
\end{thebibliography}
\end{document}